\documentclass[iop,apj]{emulateapj}
\usepackage{graphicx,url}

\shorttitle{Segue 2: The Least Massive Galaxy}
\shortauthors{Kirby et al.}
\slugcomment{Accepted for publication in ApJ on April 21, 2013}

\newcommand{\nrepeat}{35}

\newcommand{\vsyserrmeas}{1.95}
\newcommand{\meanv}{-40.2}
\newcommand{\meanverr}{0.9}

\newcommand{\sigmavlimn}{2.2}
\newcommand{\sigmavlimnf}{2.6}

\newcommand{\mstar}{1000}

\newcommand{\masslimn}{1.5}
\newcommand{\masslimnf}{2.1}

\newcommand{\mllimn}{ 360}
\newcommand{\mllimnf}{ 500}

\newcommand{\nfeh}{287}
\newcommand{\nfehmember}{10}
\newcommand{\nalphamember}{8}
\newcommand{\fehmin}{-2.85}

\newcommand{\fehmax}{-1.33}
\newcommand{\fehmean}{-2.22}
\newcommand{\fehsigma}{0.43}
\newcommand{\fehmeanerr}{0.13}
\newcommand{\lzrfeh}{-2.83}
\newcommand{\lzrfehwilone}{-2.81}
\newcommand{\lzrfehsegone}{-2.98}
\newcommand{\clzrchance}{10}
\newcommand{\clzrchancedev}{1786}
\newcommand{\lzrchancedev}{0.2}

\newcommand{\ntargets}{647}
\newcommand{\ngood}{349}
\newcommand{\nmember}{25}

\newcommand{\nbesanconcontaminantsraw}{276}
\newcommand{\nbesanconcontaminants}{0.9}

\newcommand{\nbelmatch}{13}

\newcommand{\oddballname}{SDSS J021904.93+200715.4}
\newcommand{\oddballfeh}{-2.29}
\newcommand{\oddballfeherr}{0.12}
\newcommand{\oddballfehcat}{-3.23}
\newcommand{\oddballfehcaterr}{0.05}
\newcommand{\rocheradius}{1.1}
\newcommand{\masstide}{80}
\newcommand{\distgc}{41}
\newcommand{\rrlname}{SDSS J021900.06+200635.2}
\newcommand{\closestnonmemname}{SDSS J021904.48+200218.4}
\newcommand{\closestnonmemdist}{8.6}
\newcommand{\closestnonmemrh}{2.5}

\newcommand{\closestnonmemsigma}{5.8}

\begin{document}

\title{Segue 2: The Least Massive Galaxy\altaffilmark{*}}

\author{Evan~N.~Kirby\altaffilmark{1,2},
  Michael Boylan-Kolchin\altaffilmark{1,2},
  Judith~G.~Cohen\altaffilmark{3}, 
  Marla Geha\altaffilmark{4},
  James S. Bullock\altaffilmark{1},
  and Manoj Kaplinghat\altaffilmark{1}}

\altaffiltext{*}{The data presented herein were obtained at the
  W.~M.~Keck Observatory, which is operated as a scientific
  partnership among the California Institute of Technology, the
  University of California and the National Aeronautics and Space
  Administration. The Observatory was made possible by the generous
  financial support of the W.~M.~Keck Foundation.}
\altaffiltext{1}{University of California, Department of Physics and
  Astronomy, 4129 Frederick Reines Hall, Irvine, CA 92697, USA}
\altaffiltext{2}{Center for Galaxy Evolution Fellow.}
\altaffiltext{3}{California Institute of Technology, Department of
  Astronomy \& Astrophysics, 1200 E.\ California Blvd., MC 249-17,
  Pasadena, CA 91125, USA}
\altaffiltext{4}{Yale University, Department of Astronomy, 260 Whitney
  Ave., New Haven, CT 06511, USA}



\begin{abstract}

Segue~2, discovered by \citet{bel09}, is a galaxy with a luminosity of
only $900~L_{\sun}$.  We present Keck/DEIMOS spectroscopy of
\nmember\ members of Segue~2---a threefold increase in spectroscopic
sample size.  The velocity dispersion is too small to be measured with
our data.  The upper limit with 90\% (95\%) confidence is $\sigma_v <
\sigmavlimn$ $(\sigmavlimnf)$~km~s$^{-1}$, the most stringent limit
for any galaxy.  The corresponding limit on the mass within the 3-D
half-light radius (46~pc) is $M_{1/2} < \masslimn$ $(\masslimnf)
\times 10^5~M_{\sun}$.  Segue~2 is the least massive galaxy known.  We
identify Segue~2 as a galaxy rather than a star cluster based on the
wide dispersion in [Fe/H] (from $\fehmin$ to $\fehmax$) among the
member stars.  The stars' [$\alpha$/Fe] ratios decline with increasing
[Fe/H], indicating that Segue~2 retained Type~Ia supernova ejecta
despite its presently small mass and that star formation lasted for at
least 100~Myr.  The mean metallicity, $\langle \rm{[Fe/H]} \rangle =
\fehmean \pm \fehmeanerr$ (about the same as the Ursa Minor galaxy,
330 times more luminous than Segue~2), is higher than expected from
the luminosity--metallicity relation defined by more luminous dwarf
galaxy satellites of the Milky Way.  Segue~2 may be the barest remnant
of a tidally stripped, Ursa Minor-sized galaxy.  If so, it is the best
example of an ultra-faint dwarf galaxy that came to be ultra-faint
through tidal stripping.  Alternatively, Segue~2 could have been born
in a very low-mass dark matter subhalo ($v_{\rm max} <
10$~km~s$^{-1}$), below the atomic hydrogen cooling limit.

\end{abstract}


\section{Introduction}
\label{sec:intro}

The Sloan Digital Sky Survey \citep[SDSS,][]{aba09} has revolutionized
the concept of a galaxy.  The sky coverage and depth of SDSS enabled
the discovery of low-luminosity, low-surface brightness galaxies.
Because their low surface brightnesses limit the distance out to which
SDSS can detect them, almost all of the new SDSS dwarfs are satellites
of the Milky Way (MW)\@.  The most luminous of the new satellites,
Canes Venatici~I, has an absolute magnitude of $M_V = -8.6$ \citep[$L
  = 2.3 \times 10^5~L_{\sun}$,][]{zuc06,mar08}, which overlaps the
lower luminosity bound of the previously known satellites.  The least
luminous new satellite is Segue~1, with $M_V = -1.5$ \citep[$L =
  340~L_{\sun}$,][]{bel07,mar08,geh09,sim11}.

The extremely low luminosities and stellar masses of these galaxies
prompted \citet{wil12} to suggest a new definition of a galaxy to
distinguish it from a star cluster.  Resolved stellar spectroscopy of
all of the ultra-faint ($L < 10^5~L_{\sun}$) satellites known in
\citeyear{sim07} revealed stellar velocity dispersions far in excess
of the level that would be anticipated from their stellar masses alone
\citep[e.g.,][]{sim07}.  Therefore, \citeauthor{wil12} defined a
galaxy as ``a gravitationally bound collection of stars whose
properties cannot be explained by a combination of baryons and
Newton’s laws of gravity.''  The definition is phrased to include
cosmologies involving dark matter or post-Newtonian modifications to
gravity.  The definition also does not mandate \textit{kinematic}
confirmation of dark matter-like phenomenology in order to classify a
star system as a galaxy.  Evidence for the retention of supernova
ejecta beyond what would be possible from the current baryonic mass
alone can also suffice.  A dispersion in heavy elements serves as
proof of supernova self-enrichment and therefore as confirmation of a
galaxy.  The supernova self-enrichment test is important for the
present work.

The low stellar densities of the SDSS ultra-faint dwarf galaxies make
it difficult to detect them.  These galaxies are near enough that
their stellar populations are resolved.  However, they contain so few
stars that foreground stars and background unresolved galaxies
outnumber the dwarf galaxy's own stars.  The galaxy's stars are found
instead by using matched filters, wherein only stars with appropriate
colors and magnitudes are considered possible members
\citep{roc02,wal09}.  As SDSS photometric catalogs became publicly
available over the last decade, a flurry of new MW satellite galaxies
were discovered using matched filters.

The first generation of SDSS (SDSS-I) has been exhausted as a
discovery survey for new satellites.  Additional satellites may be
found by deeper surveys or by surveys that target parts of the 75\% of
the sky that SDSS-I did not observe.  One such survey was the Sloan
Extension for Galactic Understanding and Evolution
\citep[SEGUE,][]{yan09}, which was part of SDSS-II\@.  SEGUE
discovered three additional satellites: the previously mentioned
Segue~1, the ultra-faint galaxy Segue~2 \citep[][hereafter
  B09]{bel09}, and the globular cluster Segue~3.  Segue~3 is known to
be a globular cluster because it has neither kinematic evidence of
dark matter nor a dispersion of metallicity \citep{fad11}.  Segue~2 is
the subject of this article.

\citeauthor*{bel09} discovered Segue~2 (also called the Aries
ultra-faint dwarf) in SEGUE imaging.  They obtained spectroscopy and
deeper imaging with the Hectospec and Megacam instruments on the
MMT\@.  They found that Segue~2 has a luminosity of $M_V = -2.5$
($900~L_{\sun}$).  From the individual radial velocities of five red
giants, they measured a velocity dispersion of
$3.4_{-1.2}^{+2.5}$~km~s$^{-1}$.  The expected velocity dispersion in
the absence of dark matter is 0.5~km~s$^{-1}$.  Hence, Segue~2 seemed
to contain significantly more dark matter than luminous matter.

\citeauthor*{bel09} also found tentative evidence for a stellar stream
at the same position as Segue~2.  It was detected as an overdensity of
stars at the same radial velocity as Segue~2, but with larger
metallicities than the gravitationally bound stars.  The stream
occupies a larger area of sky than the galaxy.  The presence of the
stream is exciting because it could be that Segue~2 was deposited in
the MW halo as a satellite or subcomponent of a larger galaxy that has
been tidally disrupted.  A similar origin has been proposed for
stellar systems that may have arrived via the tidal dissolution of the
Sagittarius dwarf galaxy \citep[e.g.,][]{law10}.

One of the most interesting properties of Segue~2 is its average
metallicity.  Dwarf spheroidal galaxies (dSphs) obey a universal
relationship between average stellar metallicity and luminosity or
stellar mass \citep{kir11b}.  The relation predicts that a galaxy of
Segue~2's luminosity should have $\langle {\rm [Fe/H]} \rangle =
\lzrfeh$.  From the coadded spectrum of five red clump giants,
\citeauthor*{bel09} determined their average metallicity to be
$\langle {\rm [Fe/H]} \rangle = -2.0 \pm 0.25$.  Other MW satellites
about as faint as Segue~2 also lie above the luminosity--metallicity
relation (LZR)\@.  Segue~1 should have an average metallicity of
$\langle {\rm [Fe/H]} \rangle = \lzrfehsegone$ based on the LZR, but
\citet{sim11} measured $\langle {\rm [Fe/H]} \rangle = -2.38 \pm
0.37$, and \citet{var13} measured $\langle {\rm [Fe/H]} \rangle =
-2.03 \pm 0.06$.  Willman~1 should have $\langle {\rm [Fe/H]} \rangle
= \lzrfehwilone$, but \citet{wil11} measured $\langle {\rm [Fe/H]}
\rangle = -2.19 \pm 0.46$.  The deviations from the tight relationship
between metallicity and luminosity might indicate that the faintest MW
satellites are tidally stripped remnants of galaxies that were once
hundreds of times more luminous.  Or they could hint at a metallicity
floor for galaxy evolution.  It is important to measure metallicities
for individual stars in Segue~2 to test whether the average
metallicity for a larger sample remains above the LZR\@.

We observed stars in and around Segue~2 in order to enlarge the
available spectroscopic sample, to refine the velocity dispersion and
metallicities, and to measure detailed abundances.  The main purpose
of our study is to suggest possible formation and evolution mechanisms
for Segue~2.  How long did it take to form stars?  Why does it have so
few stars today?  Was it always so faint?  In Section~\ref{sec:obs},
we describe our observations.  In Section~\ref{sec:spectroscopy}, we
detail our method for measuring radial velocities, chemical
abundances, and their uncertainties.  Because foreground stars
outnumber Segue~2 stars by a large factor, we carefully consider
membership in Section~\ref{sec:membership}.  From spectroscopy of the
member stars, we measure the galaxy's dynamical properties
(Section~\ref{sec:dynamics}) and chemical abundance pattern
(Section~\ref{sec:chemistry}).  Section~\ref{sec:dark_matter_physics}
discusses Segue~2's relevance to dark matter physics.  Finally,
Section~\ref{sec:summary} summarizes our methodology and findings.


\section{Spectroscopic Observations}
\label{sec:obs}

We observed individual objects in the vicinity of Segue~2 with the
DEIMOS spectrograph \citep{fab03} on the Keck~II telescope.  The
target density in this region of the sky allowed about 80
spectroscopic targets per slitmask.

\subsection{Target selection}
\label{sec:selection}

We placed ten slitmasks on and around the center of Segue~2.  The
slitmasks were contiguous on the sky.  We selected targets from Data
Release 7 of SDSS \citep{aba09} by querying CasJobs, the online
database tool for SDSS, for point sources classified as stars in the
30' around Segue~2.  In the instances when slitmask design constraints
forced a choice among multiple objects, we chose the object that lay
closest in $g-r$ color to a Yonsei-Yale theoretical isochrone
\citep{dem04} with an age of 14~Gyr and ${\rm [Fe/H]} = -1.5$.  We
chose ${\rm [Fe/H]} = -1.5$ rather than ${\rm [Fe/H]} = -2$
(\citeauthor*{bel09}) to be more inclusive of metal-rich stars.  For
example, a star at ${\rm [Fe/H]} = -3$ is slightly bluer than the
${\rm [Fe/H]} = -2$ isochrone, but a star at ${\rm [Fe/H]} = -1$ is
much redder.  Therefore, we chose a central isochrone more metal-rich
than the measured mean metallicity.  In practice, the sample is almost
entirely free of color or metallicity bias because the stellar density
rarely forced a choice among multiple spectroscopic targets.  In the
cases when a brighter star lay about as close to the isochrone as a
fainter star, we chose the brighter star.

\begin{figure}[t!]
\centering
\includegraphics[width=\columnwidth]{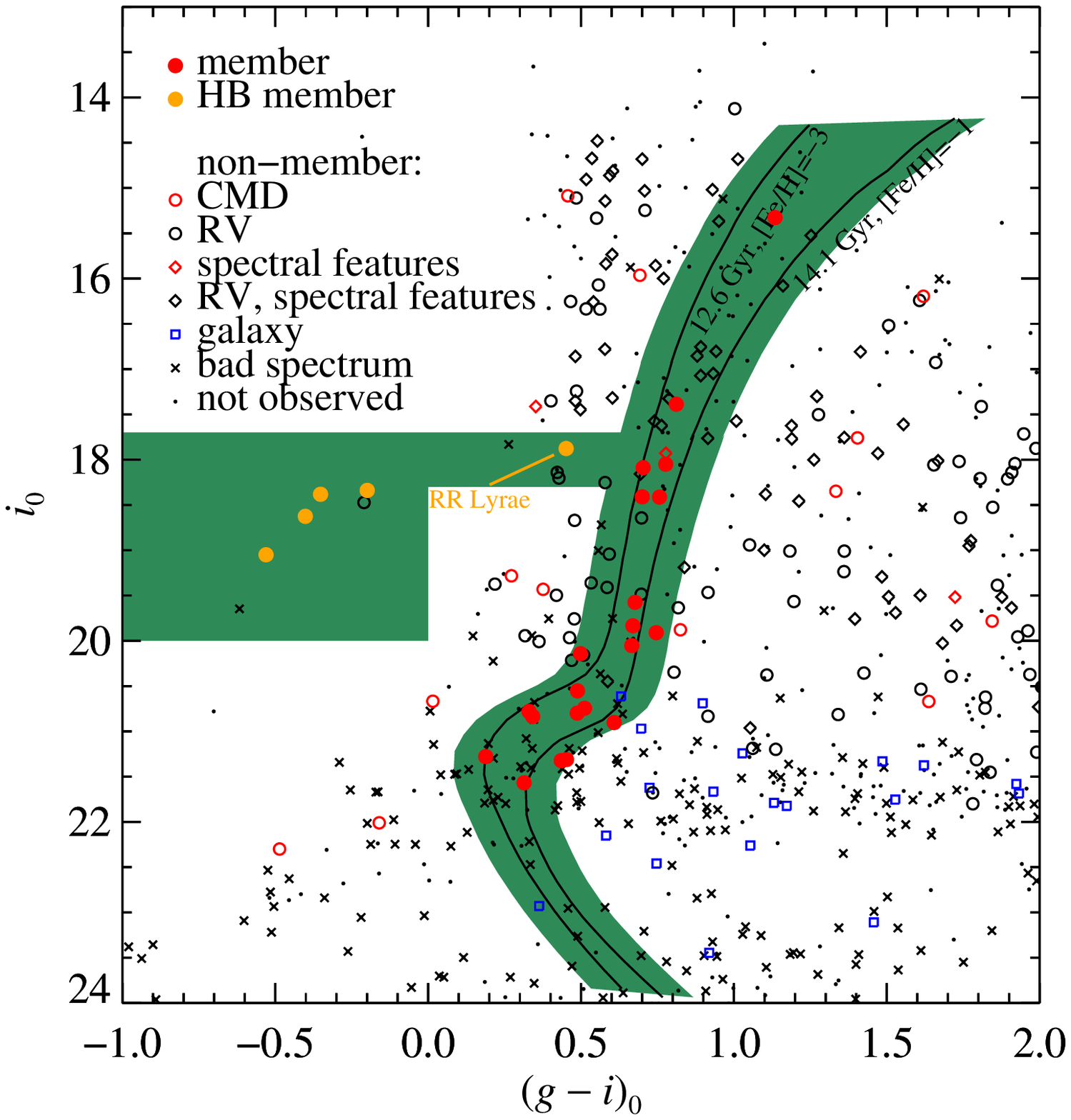}
\caption{Color-magnitude diagram from SDSS \protect \citep{aba09} for
  the 160~arcmin$^2$ of sky centered on Segue~2 (same area shown in
  Figure~\ref{fig:map}).  The figure legend lists reasons for which
  stars were excluded from consideration as members.  Additionally,
  stars outside of the shaded region, which is based on Yonsei-Yale
  theoretical isochrones \protect \citep{dem04}, are ruled
  non-members.  The black curves show isochrones of two different ages
  and metallicities.  Filled circles are member stars.  Red/orange
  (black) symbols identify stars that passed (failed) the radial
  velocity test.  Diamonds indicate stars that show spectral features,
  such as a strong \ion{Na}{1}~8190 doublet, indicating that they are
  foreground dwarfs.  Blue squares indicate spectroscopically
  identified galaxies.\label{fig:cmd}}
\end{figure}

\begin{figure}[t!]
\centering
\includegraphics[width=\columnwidth]{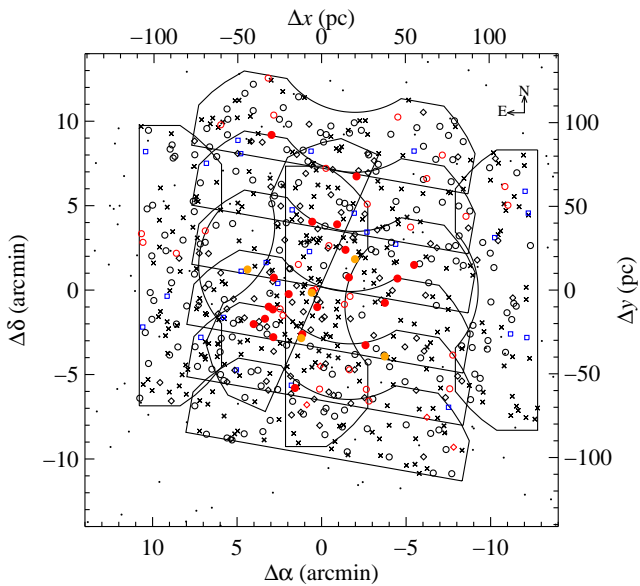}
\caption{Sky map centered on Segue~2 \protect ($\alpha_0 =
  2^{\rm{h}}19^{\rm{m}}16^{\rm{s}}$, $\delta_0 = +20\arcdeg 10\arcmin
  31\arcsec$, \citeauthor*{bel09}).  The outlines of the ten DEIMOS
  slitmasks enclose the objects observed.  The symbols have the same
  meanings as in Figure~\ref{fig:cmd}.\label{fig:map}}
\end{figure}

Figure~\ref{fig:cmd} shows a color-magnitude diagram for the region of
the sky enclosed by the DEIMOS slitmasks.  Figure~\ref{fig:map} shows
the slitmask placement on the sky.  The area of sky covered is about
0.12~deg$^2$, which is about one sixth of the area covered by
\citeauthor*{bel09}'s Hectospec survey.  In both figures, all symbols
other than the smallest dots indicate targets for which we obtained
spectra.  The scarcity of the smallest dots in these figures shows
that we observed many of the possible member red giants, subgiants,
and horizontal branch (HB) stars in Segue~2.  There were
\ntargets\ unique spectroscopic targets.  Of these, \ngood\ were stars
with spectral quality sufficient to recover a radial velocity.  Only
\nmember\ of these stars are likely members of Segue~2 (see
Section~\ref{sec:membership}).

One star (\rrlname) on the HB turned out to be an RR Lyrae star.
\citet{boe13} discovered the star's variability with multi-epoch
photometry of Segue~2.  Because RR Lyrae stars vary in their observed
radial velocity by 50--70~km~s$^{-1}$ even in the weak metal lines
\citep{pre64b,ses12}, we excluded \rrlname\ from the measurement of
the velocity dispersion.  We also excluded it from the measurement of
chemical abundances because the metal lines of RR Lyrae stars are too
weak for medium-resolution spectroscopy and because RR Lyrae stars are
best observed at a specific phase to obtain high-resolution
spectroscopic abundances \citep{pre64a,for11}.

\subsection{Observations}

\begin{deluxetable*}{lcr@{ }c@{ }lcclc}
\tablewidth{0pt}
\tablecolumns{9}
\tablecaption{DEIMOS Observations\label{tab:obs}}
\tablehead{\colhead{Slitmask} & \colhead{\# targets} & \multicolumn{3}{c}{Date} & \colhead{Airmass} & \colhead{Seeing} & \colhead{Individual Exposures} & \colhead{Total Exposure Time} \\
\colhead{ } & \colhead{ } & \multicolumn{3}{c}{ } & \colhead{ } & \colhead{($''$)} & \colhead{(s)} & \colhead{(s)}}
\startdata
Ari-2   & 85 & 2009 & Feb & 19 & 1.29 & $0.8$ & $1200 + 1600$ & 2800 \\
seg2\_1 & 89 & 2009 & Oct & 13 & 1.22 & $0.6$ & $3 \times 1200$ & 3600 \\
seg2\_2 & 91 & 2009 & Oct & 13 & 1.03 & $0.6$ & $4 \times 1200$ & 4800 \\
seg2\_3 & 86 & 2009 & Oct & 13 & 1.22 & $0.6$ & $3 \times 1200$ & 3600 \\
seg2\_4 & 81 & 2009 & Oct & 14 & 1.50 & $0.5$ & $3 \times 1200 + 534 + 700$ & 4834 \\
seg2\_5 & 83 & 2009 & Oct & 14 & 1.08 & $0.4$ & $4 \times 1200$ & 4800 \\
seg2\_6 & 84 & 2009 & Oct & 14 & 1.03 & $0.5$ & $4 \times 1200$ & 4800 \\
seg2\_7 & 83 & 2009 & Oct & 13 & 1.20 & $0.5$ & $3 \times 1200$ & 3600 \\
seg2\_8 & 69 & 2009 & Oct & 14 & 1.18 & $0.6$ & $3 \times 1200$ & 3600 \\
seg2\_9 & 29 & 2013 & Jan & 13 & 1.03 & $1.2$ & $6 \times 1200$ & 7200 \\
\enddata
\end{deluxetable*}

We observed the slitmasks in generally excellent weather during three
nights in 2009 and one night in 2013.  Table~\ref{tab:obs} details the
exposure times for and conditions under which each slit each slitmask
was observed.

We obtained calibration exposures in the afternoons before each
observing night.  These included three quartz flat lamp exposures and
one arc lamp exposure per slitmask.  The arc lamp exposures included
simultaneous Ne, Ar, Kr, and Xe lamps.

Slitmasks were aligned with a minimum of four 4'' alignment boxes.
Seeing was measured from the FWHM of the profiles of the alignment
stars when the grating was angled for zeroth order (no spectral
dispersion).  Spectral observations were obtained with the 1200 line
mm$^{-1}$ grating in first order.  The resolution in this
configuration was 1.2~\AA\ FWHM, corresponding to a resolving power of
$R = 7000$ at the \ion{Ca}{2} infrared triplet.  The grating has a
blaze wavelength of 7760~\AA, and we set the central wavelength to
7800~\AA\@.  Vignetting near the edges of the field of view and the
locations of slits along the dispersion axis caused variation in the
wavelength coverage from slit to slit by up to 300~\AA\@.  The typical
wavelength range for a single object was 2700~\AA.  During spectral
observations, DEIMOS's active flexure compensation system kept the
spectra stationary on the detector within a precision of 0.07~\AA\ in
the dispersion direction and $0 \farcs 02$ in the spatial direction.

\subsection{Reductions}

We reduced the raw images into one-dimensional spectra with the
\texttt{spec2d} software \citep{coo12} developed by the Deep
Extragalactic Evolutionary Probe~2 team \citep[DEEP2,][]{dav03,new13}.
Small modifications to the pipeline allowed for better extraction of
1-D spectra by treating the stars as point sources rather than
extended galaxies.  For a slightly more detailed description of the
data reduction procedure, see Section~2.3 of \citet{kir12}, who used
DEIMOS in the same configuration.

\begin{figure}[t!]
\centering
\includegraphics[width=\columnwidth]{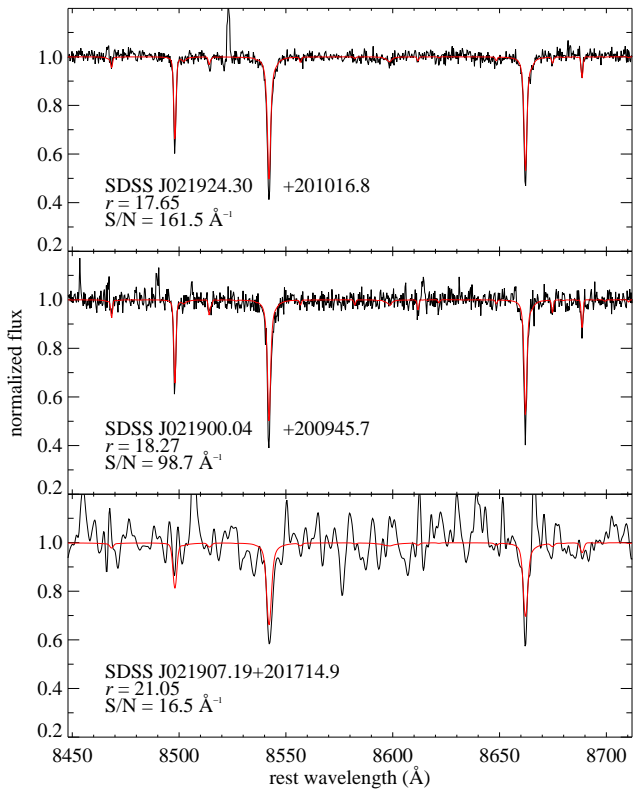}
\caption{Examples of small regions of DEIMOS spectra of three member
  stars of high, medium, and low S/N\@.  In the top two panels, the
  best-fitting synthetic spectra are plotted in red.  The spectral
  fitting for chemical abundances excludes the poorly modeled Ca
  triplet.  Because the S/N of the spectrum in the bottom panel is too
  low to fit a synthetic spectrum, the red line shows a synthetic
  spectrum with parameters representative of a subgiant in Segue~2.
  The spectra in the bottom panel are smoothed with a Gaussian kernel
  of 1.6~\AA\ FWHM.\label{fig:examples}}
\end{figure}

Figure~\ref{fig:examples} shows three examples of reduced 1-D spectra
at a variety of signal-to-noise ratios (S/Ns).  All three stars were
determined to be members (Section~\ref{sec:membership}).  The quality
of the top two spectra is high enough to measure both radial
velocities (Section~\ref{sec:rv}) and chemical abundances
(Section~\ref{sec:abundances}).  The quality of the bottom spectrum is
good enough to measure its radial velocity but not metallicity or
chemical abundance ratios to the required precision.


\section{Spectroscopic Measurements}
\label{sec:spectroscopy}

From the reduced 1-D spectra, we measured stellar radial velocities
and chemical abundances.  The procedures for both measurements closely
mimic those described by \citet{kir12}.  Please refer to that article
for details not included here.

\subsection{Radial Velocity Measurements}
\label{sec:rv}

We cross-correlated each 1-D spectrum with template spectra obtained
by \citet{sim07}.  We adopted the radial velocity ($v_r$) at the
cross-correlation peak of the template spectrum with the lowest
$\chi^2$ when compared to the observed spectrum.  All of the radial
velocities were checked by visually comparing the template spectrum to
the velocity-corrected observed spectrum.  Because astrometric
uncertainty and imperfect slitmask alignment can cause a star to be
mis-centered in its DEIMOS slitlet, we also applied a correction to
$v_r$ based on the observed wavelengths of telluric absorption bands.
\citet{soh07} first applied this technique to DEIMOS spectra.  For
each DEIMOS spectrum, we cross-correlated the observed A and B
molecular absorption bands imprinted by the Earth's atmosphere with
the template spectrum of a hot star.  We then applied a correction to
$v_r$ to align the telluric bands of each observed spectrum with the
template spectrum.  Finally, we shifted all velocities to the
heliocentric frame.  All velocities quoted in this article are
heliocentric velocities ($v_{\rm helio}$).

We estimated uncertainties on $v_r$ by resampling the spectrum and
repeating the cross-correlation.  The \texttt{spec2d} reduction
software produced an estimate of the noise in each pixel.  We added
Gaussian random noise to each spectrum based on this array.  This
process degraded the S/N by $\sqrt{2}$.  We measured $v_r$ from the
noise-added array.  Then, we resampled the spectrum with a different
Gaussian random noise array.  In all, we measured $v_r$ from 1000
realizations of the spectrum.  The random radial velocity uncertainty
from the Monte Carlo resampling, $\delta_{\rm MC} v_r$, is the
standard deviation of the 1000 measurements of $v_r$.

\citet{sim07} found that $\delta_{\rm MC} v_r$ is an incomplete
description of the error on $v_r$.  In particular, they found a
systematic error floor of $\delta_{\rm sys} v_r = 2.2$~km~s$^{-1}$.
The systematic error is added in quadrature with the random
uncertainty such that the total error on $v_r$ is $\delta v_r =
\sqrt{\delta_{\rm MC}v_r^2 + \delta_{\rm sys}v_r^2}$.
\citeauthor{sim07}\ estimated this error from 49 repeat measurements
of $v_r$ for the same stars on different DEIMOS slitmasks.  From a
different set of 106 repeat observations in the galaxy VV124,
\citet{kir12} calculated $\delta_{\rm sys} v_r = 2.21$~km~s$^{-1}$.
Our data set for Segue~2 includes \nrepeat\ measurements that satisfy
all membership cuts (Section~\ref{sec:membership}) except radial
velocity.  From these \nrepeat\ stars, we calculated $\delta_{\rm sys}
v_r = \vsyserrmeas$~km~s$^{-1}$.  Because this estimate of systematic
error was determined from the same data set, we used it for the
remainder of our analysis.  We also repeated our analysis adopting
$\delta_{\rm sys} v_r = 2.2$~km~s$^{-1}$.  The slightly larger
systematic error did not change any of our qualitative conclusions,
and it changed the quantitative limits we placed on the velocity
dispersion and mass by only a few percent.

Measuring $v_r$ was not possible for many of the spectroscopic
targets.  The most common failure mode was low S/N, which prevented
the identification of a clear cross-correlation peak.  All spectra
were compared to a template spectrum shifted to the radial velocity of
the target star in order to verify that the cross-correlation
succeeded.  In the event that none of H$\alpha$ or the three strong
\ion{Ca}{2} triplet lines at 8498, 8542, or 8662~\AA\ could be
recognized, the spectrum was flagged as ``Bad,'' meaning that it was
of insufficient quality to measure $v_r$ confidently.  Some spectra
also suffered reduction problems, where large portions were missing.
These spectra were also flagged as Bad.  Some targets turned out to be
galaxies (see Section~\ref{sec:membershipspec}), and we did not
attempt to measure their redshifts.

In order to maximize the spectral S/N and minimize measurement errors
for the 47 stars observed on multiple slitmasks, we coadded the
individual spectra.  The individual spectra were shifted into the
same heliocentric frame before coaddition.  The coaddition weighted
each pixel by its inverse variance so that the S/N of the coadded
spectrum was maximized.  We measured radial velocities and chemical
abundances from the coadded spectra.

\addtocounter{table}{1}

Table~\ref{tab:catalog} lists all of the spectroscopic targets and,
where possible, the measurements of $v_{\rm helio}$ and their
uncertainties.  The table also includes SDSS identifications and
photometry.

\subsubsection{Comparison to \citeauthor*{bel09}}

\begin{figure}[t!]
\centering
\includegraphics[width=\columnwidth]{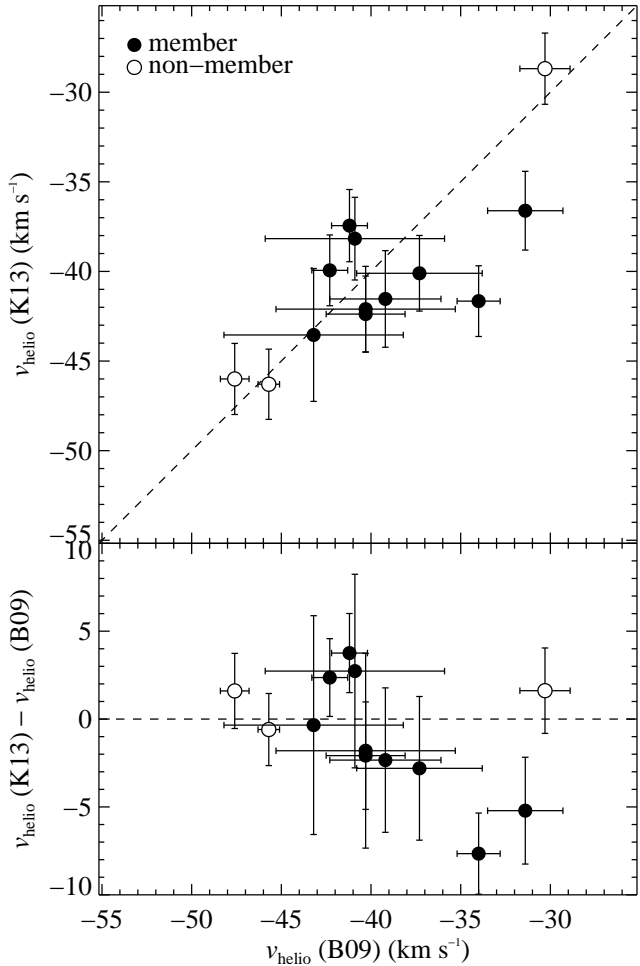}
\caption{Comparison of radial velocities from this work to
  \citeauthor*{bel09}.  The velocities of all of the stars agree to
  within the combined $2\sigma$ errors.  Of the 13 stars, 9 agree to
  within the $1\sigma$ errors.  \citeauthor*{bel09} classified the
  three non-member stars as stream members.\label{fig:bel09}}
\end{figure}

\begin{deluxetable*}{llcccccc}
\tablewidth{0pt}
\tablecolumns{8}
\tablecaption{Comparison of Radial Velocities with \protect \citet{bel09}\label{tab:bel09}}
\tablehead{ \colhead{ } & \colhead{ } & \multicolumn{2}{c}{$v_r$} & \colhead{ } & \colhead{ } & \multicolumn{2}{c}{Member?} \\ \cline{3-4} \cline{7-8}
\colhead{ID (SDSS)} & \colhead{ID (\citeauthor*{bel09})} & \colhead{K13} & \colhead{B09} & \colhead{$v_r~({\rm K13}) - v_r~({\rm B09})$} & \colhead{$\sigma$} & \colhead{K13\tablenotemark{a}} & \colhead{B09\tablenotemark{a,b}} \\
\colhead{ } & \colhead{ } & (km~s$^{-1}$) & (km~s$^{-1}$) & (km~s$^{-1}$) & \colhead{ } & \colhead{ } & \colhead{ }}
\startdata
SDSS J021904.93+200715.4 & Seg2-023 & $ -40.1 \pm  2.1$ & $ -37.3 \pm  3.5$ & $  -2.8 \pm  4.1$ & $-0.7$ & Y & Y \\
SDSS J021909.97+201254.0 & Seg2-021 & $ -42.4 \pm  2.1$ & $ -40.3 \pm  2.2$ & $  -2.1 \pm  3.1$ & $-0.7$ & Y & ? \\
SDSS J021904.38+201837.4 & Seg2-063 & $ -46.3 \pm  2.0$ & $ -45.7 \pm  0.6$ & $  -0.6 \pm  2.1$ & $-0.3$ & N & S \\
SDSS J021900.04+200945.7 & Seg2-024 & $ -37.4 \pm  2.0$ & $ -41.2 \pm  1.0$ & $  +3.8 \pm  2.2$ & $+1.7$ & Y & Y \\
SDSS J021920.87+200754.0 & Seg2-007 & $ -36.6 \pm  2.2$ & $ -31.4 \pm  2.1$ & $  -5.2 \pm  3.0$ & $-1.7$ & Y & ? \\
SDSS J021836.75+201217.8 & Seg2-069 & $ -46.0 \pm  2.0$ & $ -47.6 \pm  0.8$ & $  +1.6 \pm  2.1$ & $+0.7$ & N & S \\
SDSS J021918.49+201021.9 & Seg2-003 & $ -42.1 \pm  2.4$ & $ -40.3 \pm  5.0$ & $  -1.8 \pm  5.5$ & $-0.3$ & B & B \\
SDSS J021904.48+200218.4 & Seg2-056 & $ -28.7 \pm  2.0$ & $ -30.3 \pm  1.4$ & $  +1.6 \pm  2.4$ & $+0.7$ & N & S \\
SDSS J021922.71+200443.3 & Seg2-033 & $ -41.7 \pm  2.0$ & $ -34.0 \pm  1.2$ & $  -7.7 \pm  2.3$ & $-3.3$ & Y & Y \\
SDSS J021934.68+201144.3 & Seg2-029 & $ -43.5 \pm  3.7$ & $ -43.2 \pm  5.0$ & $  -0.3 \pm  6.2$ & $-0.1$ & B & B \\
SDSS J021929.33+200931.9 & Seg2-016 & $ -41.5 \pm  2.7$ & $ -39.2 \pm  3.1$ & $  -2.3 \pm  4.1$ & $-0.6$ & Y & ? \\
SDSS J021907.59+201220.8 & Seg2-011 & $ -38.2 \pm  2.3$ & $ -40.9 \pm  5.0$ & $  +2.7 \pm  5.5$ & $+0.5$ & B & B \\
SDSS J021924.30+201016.8 & Seg2-006 & $ -39.9 \pm  2.0$ & $ -42.3 \pm  1.0$ & $  +2.4 \pm  2.2$ & $+1.1$ & Y & Y \\
\enddata
\tablenotetext{a}{B denotes horizontal branch stars.}
\tablenotetext{b}{S denotes stars that \citeauthor*{bel09} classified as members of a stellar stream coincident with Segue~2.  \citeauthor*{bel09} considered some red giants just above the subgiant branch to be possible members, and they are denoted by a question mark (?).}
\end{deluxetable*}

Our sample overlaps with \nbelmatch\ stars in \citeauthor*{bel09}'s
sample.  Figure~\ref{fig:bel09} shows the comparison between our
measurements of $v_{\rm helio}$ and theirs.  All but four stars agree
to within the combined $1\sigma$ uncertainties, and three of those
four stars agree to within the $2\sigma$ uncertainties.
Table~\ref{tab:bel09} also shows the comparison between our work (K13)
and \citeauthor*{bel09}.

\subsection{\ion{Na}{1}~8190 Doublet}

The major contaminants in the Segue~2 spectroscopic sample are
foreground MW dwarf stars.  These stars have large surface gravities,
which are reflected in the strength of some absorption lines.  In the
DEIMOS spectral range, the lines most affected by surface gravity are
the \ion{Na}{1} doublet at 8190~\AA\@.  Although the doublet is in a
spectral region affected by telluric absorption, the equivalent width
(EW) in a typical dwarf star is 1~\AA, strong enough to be noticed
easily through the telluric absorption.

Most spectroscopic targets did not have a detectable \ion{Na}{1}
doublet.  For the rest, we measured the EW of each line in the doublet
by fitting a Gaussian or a Lorentzian profile, depending on which
profile better matched the observed line.  The sum of the two EWs is
EW(Na), which is given in Table~\ref{tab:catalog}.  These measurements
were used to rule some stars as non-members
(Section~\ref{sec:membershipspec}).

\subsection{Chemical Abundance Measurements}
\label{sec:abundances}

\citet{kir08} showed that the resolution of DEIMOS spectra hardly
limits the ability to measure iron abundances in red giants compared
to high-resolution spectroscopy.  Later, \citet{kir10} demonstrated
that [Fe/H] can be measured to a precision of 0.11~dex for DEIMOS
spectra with high S/N\@.  Furthermore, [Mg/Fe], [Si/Fe], [Ca/Fe], and
[Ti/Fe] can be measured with comparable precision.

The abundances were measured by a $\chi^2$ comparison of the observed
spectrum to a grid of synthetic spectra.  Only neutral atomic
absorption lines were used.  The \ion{Ca}{2} triplet was specifically
excluded.  In the initial stages of the measurement, the effective
temperature, surface gravity, and metallicity were based on
photometry.  In the case of Segue~2, we used 14~Gyr Padova isochrones
\citep{gir02}, which are available in the SDSS filters.  In successive
iterations, the temperature and metallicity were refined by fitting
small spectral regions around \ion{Fe}{1} lines.  In the last steps,
the four abundance ratios mentioned above were measured by restricting
the spectral matching separately to neutral Mg, Si, Ca, and Ti lines.

\begin{deluxetable*}{lccccc}
\tablewidth{0pt}
\tablecolumns{6}
\tablecaption{Abundances of Member Stars\label{tab:abundances}}
\tablehead{\colhead{ID} & \colhead{[Fe/H]} & \colhead{[Mg/Fe]} & \colhead{[Si/Fe]} & \colhead{[Ca/Fe]} & \colhead{[Ti/Fe]}}
\startdata
SDSS J021904.93+200715.4 & $-2.29 \pm 0.12$ & $+0.71 \pm 0.34$ & $+0.09 \pm 0.37$ &     \nodata      &     \nodata      \\
SDSS J021920.87+200754.0 & $-2.30 \pm 0.18$ &     \nodata      & $+0.52 \pm 0.40$ & $+0.45 \pm 0.38$ &     \nodata      \\
SDSS J021922.71+200443.3 & $-2.25 \pm 0.11$ &     \nodata      & $+0.46 \pm 0.20$ & $+0.30 \pm 0.17$ & $+0.40 \pm 0.22$ \\
SDSS J021909.97+201254.0 & $-1.33 \pm 0.12$ & $+0.10 \pm 0.46$ & $-0.25 \pm 0.33$ &     \nodata      & $-0.26 \pm 0.43$ \\
SDSS J021917.10+200930.6 & $-2.20 \pm 0.13$ &     \nodata      & $+0.62 \pm 0.17$ & $+0.34 \pm 0.21$ & $+0.70 \pm 0.18$ \\
SDSS J021924.30+201016.8 & $-2.53 \pm 0.12$ &     \nodata      & $+0.80 \pm 0.14$ & $+0.78 \pm 0.14$ & $+0.67 \pm 0.15$ \\
SDSS J021900.04+200945.7 & $-1.91 \pm 0.11$ &     \nodata      & $-0.05 \pm 0.23$ & $+0.27 \pm 0.19$ & $-0.10 \pm 0.27$ \\
SDSS J021929.33+200931.9 & $-2.68 \pm 0.31$ &     \nodata      &     \nodata      &     \nodata      &     \nodata      \\
SDSS J021928.04+201115.2 & $-2.03 \pm 0.32$ &     \nodata      &     \nodata      &     \nodata      &     \nodata      \\
SDSS J021933.13+200830.2 & $-2.85 \pm 0.11$ & $+0.76 \pm 0.25$ & $+0.17 \pm 0.25$ & $+0.30 \pm 0.13$ & $-0.18 \pm 0.31$ \\
\enddata
\end{deluxetable*}

We were able to measure [Fe/H] for \nfeh\ stars, assuming that they
are all at the distance of Segue~2.  At least one [$\alpha$/Fe]
abundance ratio was measurable for all but two of the \nfehmember\ of
those stars later determined to be members
(Section~\ref{sec:membership}).  Those measurements are given in
Table~\ref{tab:abundances}.

\begin{figure}[t!]
\centering
\includegraphics[width=\columnwidth]{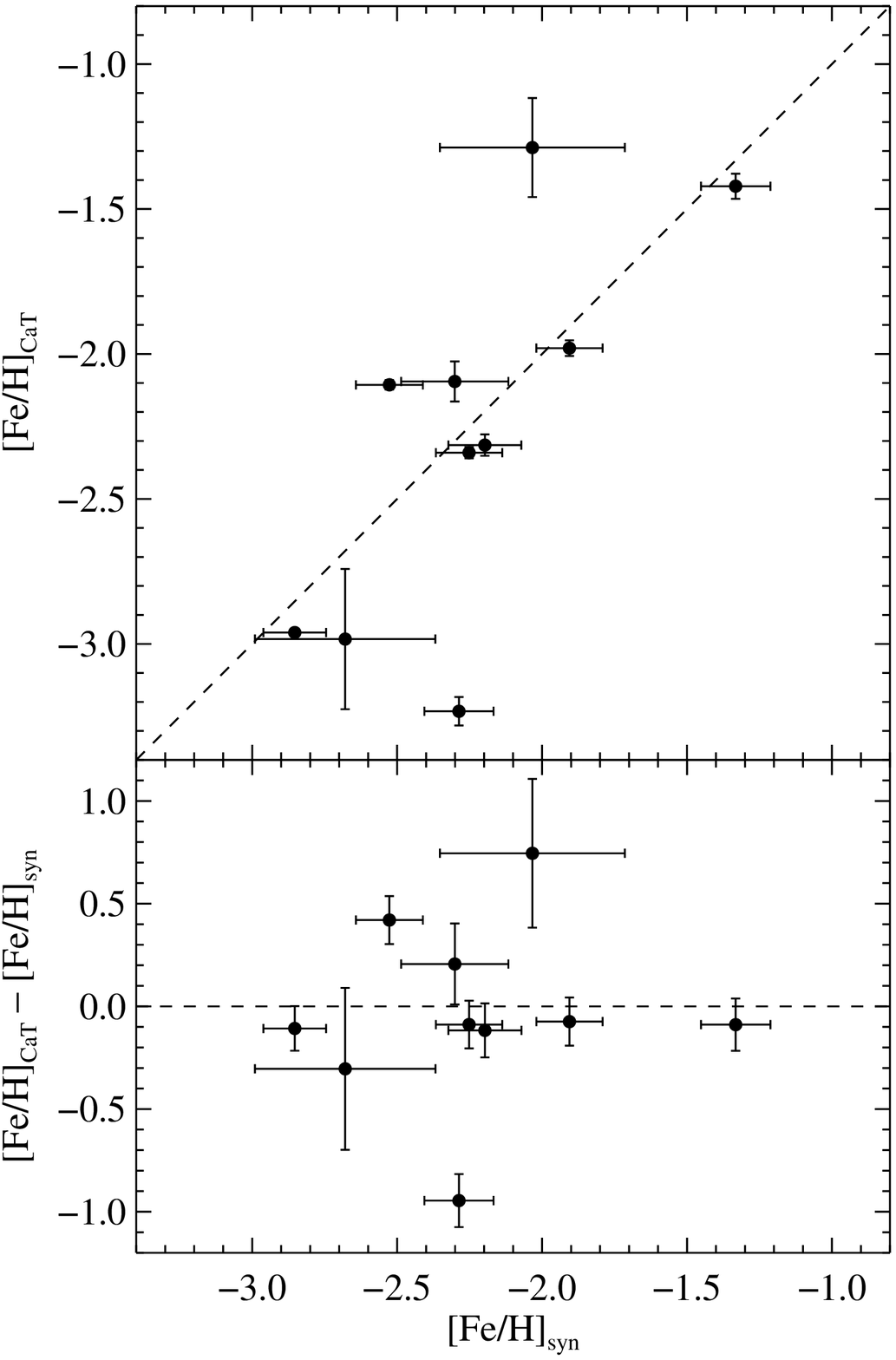}
\caption{Comparison of iron abundances based on the \ion{Ca}{2}
  triplet to iron abundances based on spectral synthesis of
  \ion{Fe}{1} lines.  All of the stars shown here are members of
  Segue~2.\label{fig:fehcat}}
\end{figure}

As a check on our [Fe/H] measurements, we also measured [Fe/H] from an
empirical calibration based on the \ion{Ca}{2} triplet \citep{sta10}.
For stars where the lines at 8542 and 8662~\AA\ were observed, we fit
separate Lorentzian profiles to each line.  We added their EWs and
applied \citeauthor{sta10}'s calibration.  This calibration requires a
$V$ magnitude, which we calculated from the SDSS magnitudes using
\citeauthor{jor06}'s (\citeyear{jor06}) metallicity-independent
transformation equations.

Figure~\ref{fig:fehcat} shows the comparison between the Ca triplet
metallicity and those from spectral synthesis of \ion{Fe}{1} lines for
those stars determined to be members by the criteria laid out in the
next section.  Six (60\%) of the measurements agree to within the
$1\sigma$ error bars.  Three more (30\%) agree to within $2\sigma$.
However, one star (\oddballname) has ${\rm [Fe/H]}_{\rm syn} =
\oddballfeh \pm \oddballfeherr$ and ${\rm [Fe/H]}_{\rm CaT} =
\oddballfehcat \pm \oddballfehcaterr$.  Visual inspection of strong
iron lines, such as \ion{Fe}{1}~8689, showed that the synthetic fit
accurately represents the Fe lines in the spectrum.  Furthermore, the
[$\alpha$/Fe] ratios measured for this star fall in line with Segue~2
stars adjacent in metallicity.  The [$\alpha$/Fe] ratio would have
been $\sim 1$~dex higher if the star truly had ${\rm [Fe/H]} =
\oddballfehcat$.  We chose to rely on the spectral synthesis
measurements rather than the Ca triplet because spectral synthesis
provides independent estimates of both [Fe/H] and [$\alpha$/Fe]
ratios.  It also measures [Fe/H] from Fe lines instead of Ca lines.


\section{Membership}
\label{sec:membership}

Because Segue~2 is such a sparse galaxy, it was necessary to remove
contaminants from the spectroscopic sample.  The primary contaminants
were foreground dwarf stars.  Evolved stars and galaxies also
contaminated the sample.  This section describes the criteria for a
star to be considered a member of Segue~2.

\subsection{Photometric Criteria}

In the absence of better candidates for membership, many objects were
placed on the DEIMOS slitmasks with the knowledge that they could not
be members of Segue~2.  These objects were spectroscopically targeted
merely to fill the slitmasks.  We ruled stars as non-members if they
lay outside of a selection area in the $i_0$ vs.\ $(g-i)_0$ CMD, shown
in Figure~\ref{fig:cmd}.  The selection area was bounded on the red
side of the red giant branch by a Yonsei-Yale isochrone with an age of
14.1~Gyr and ${\rm [Fe/H]} = -1$.  The blue bound was a Yonsei-Yale
isochrone with an age of 12.6~Gyr and ${\rm [Fe/H]} = -3$.  In order
to be more inclusive of a range of metallicities and ages and to
account for modeling errors in the isochrones, an additional buffer of
0.1~mag in color was allowed beyond the blue and red isochrones.
Finally, we also allowed stars on the HB in two selection boxes
defined by $-1.0 < (g-i)_0 < 0.0$ and $17.7 < i_0 < 20.0$ and $0.0 \le
(g-i)_0 < 0.6$ and $17.3 < i_0 < 18.3$.

\subsection{Spectroscopic Criteria}
\label{sec:membershipspec}

The first membership cut based on spectroscopy excluded background
galaxies.  If a target showed redshifted emission lines or Ca~H and K
absorption, we ruled it as a galaxy.  This criterion eliminated many
objects with $i_0 > 20$.

Second, we identified some foreground dwarfs from EW(Na).
\citet{kir12} computed synthetic EWs for the \ion{Na}{1}~8190 doublet,
and they found that any star with ${\rm EW(Na)} > 1$~\AA\ must be a
dwarf with $\log g > 4.5$.  Any stars with such high surface gravities
at the distance of Segue~2 would be far too faint to be included in
our spectroscopic sample.  They would be foreground contaminants.  We
adopted the same criterion, but we also added more restrictive
criteria for stars for which we measured [Fe/H]\@.  Based on
\citeauthor{kir12}'s (\citeyear{kir12}) Figure~6, we ruled out stars
with ${\rm EW(Na)} > 0.7$~\AA\ and $-2 \le {\rm [Fe/H]} < -1$.  We
also ruled out stars with ${\rm EW(Na)} > 0.4$~\AA\ and ${\rm [Fe/H]}
< -2$.

Next, we imposed a metallicity cut.  \citeauthor*{bel09} found
tentative evidence for a stellar stream coincident both in position
and $v_r$ with Segue~2.  From the EW of the Mg~b triplet, they found
the stream to be more metal-rich than Segue~2.  We recovered a
metallicity of ${\rm [Fe/H]} > -1$ for one star that passed all
membership cuts other than metallicity.  Because this star may belong
to the stream, we excluded it from membership consideration.

\begin{figure}[t!]
\centering
\includegraphics[width=\columnwidth]{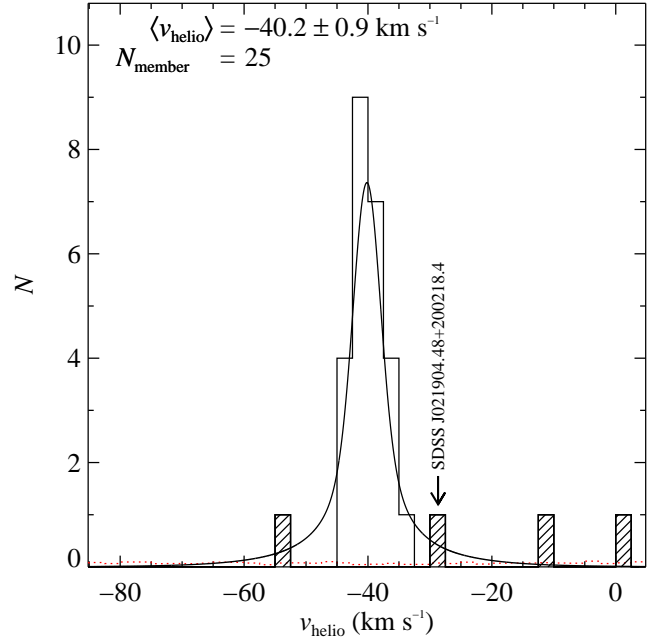}
\caption{Radial velocity distribution of stars that pass all
  membership cuts except $v_r$.  The hatched regions of the histogram
  show stars excluded from membership on the basis of $v_r$.  The red,
  dotted histogram shows the expected level of contamination of
  non-members from the Besan{\c c}on model.  The black, solid curve
  shows the expected error distribution, assuming that all stars have
  the same intrinsic radial velocity.\label{fig:vhist}}
\end{figure}

Finally, we excluded stars based on $v_{\rm helio}$.  It turned out
that our data cannot resolve the velocity dispersion of Segue~2.  In
other words, the velocity dispersion, $\sigma_v$, is consistent with
zero, such that all stars have the same intrinsic radial velocity
within the measurement uncertainties.  Therefore, our membership
criterion based on velocity did not depend on $\sigma_v$.  Instead, we
accepted all stars with $|v_{\rm helio} - \langle v_{\rm helio}
\rangle| < 2.58\delta v_r$, where $\langle v_{\rm helio} \rangle$ is
the average radial velocity, determined in Section~\ref{sec:dynamics}.
For Gaussian distributed errors, a membership cut of $2.58\delta v_r$
rejects non-members in addition to 1\% of members.  It is unlikely
that we rejected any members with the radial velocity cut because our
sample has only \nmember\ members.  Figure~\ref{fig:vhist} shows the
velocity distribution for stars that passed all membership cuts except
radial velocity.  Those stars that failed the radial velocity cut are
shaded.

One star (\closestnonmemname) excluded on the basis of radial velocity
alone had a velocity close to a member star.  It is identified in
Figure~\ref{fig:vhist}.  It was excluded because the measurement
uncertainty on $v_{\rm helio}$ was small, whereas the member star with
similar velocity had a larger measurement uncertainty.  The radial
velocity of the non-member is $\closestnonmemsigma\sigma$ discrepant
from $\langle v_{\rm helio} \rangle$, but the membership cut was
$2.58\sigma$.  If \closestnonmemname\ were counted as a member,
Segue~2's velocity dispersion would be $\sigma_v = 4.3$~km~s$^{-1}$.
We are confident that it is a non-member because its velocity is
highly discrepant from $\langle v_{\rm helio} \rangle$ and because it
is farther from the center of Segue~2 ($\closestnonmemdist'$ or
$\closestnonmemrh$ half-light radii) than all but one member star.


\subsection{Estimate of Residual Contamination}

Despite our fairly stringent membership criteria, some unidentified
contaminants might have remained in our sample.  The available
information did not allow us to further cull the sample.  Therefore,
we attempted to quantify the expected level of residual contamination.
In other words, how many stars in our ``member'' sample are likely to
be non-members?

We consulted the Besan{\c c}on model of Galactic structure
\citep{rob03}.  We used their online web
form\footnote{\url{http://model.obs-besancon.fr}} to generate a
catalog simulation.  The simulation was limited to the galactic
coordinates of the center of Segue~2.  The solid angle was 10~deg$^2$,
but the model parameters were based only on the coordinates of
Segue~2.  We chose such a large area so that the sample size was
large.  The model returned CFHT magnitudes, heliocentric radial
velocities, and metallicities for stars in the Galactic model.  The
model also added mock observational uncertainties to these quantities
according to a law that we determined empirically from our data set.
Both photometric and kinematic errors obeyed exponential forms as a
function of $i_0$ magnitude.

\begin{eqnarray}
\delta g = -0.010 + \exp(-18.665 + 0.765 i_0) \\
\delta i = -0.031 + \exp(-12.798 + 0.531 i_0) \\
\delta_{\rm MC} v_r = -0.565 + \exp(-14.730 + 0.831 i_0)
\end{eqnarray}

\noindent We added $\delta_{\rm sys} v_r$ in quadrature to
$\delta_{\rm MC} v_r$ to obtain $\delta v_r$, just as we did for the
observed velocities.  Finally, we converted CFHT to SDSS magnitudes
following \citet{reg09}.

We applied all of the membership cuts described above to the Besan{\c
  c}on model simulation.  Just as for the observational data, we
allowed model stars to pass membership cuts if colors, magnitudes, and
velocities were consistent with membership.  The metallicity cut
needed to be modified to account for the fact that we could not
recover metallicities for most faint subgiants.  Therefore, a model
star was eliminated based on metallicity only if it had ${\rm [Fe/H]}
> -1$ and $i_0 < 20.5$.  That is about the magnitude where our
estimated uncertainty on [Fe/H] exceeded 0.4~dex.  We also excluded
stars with $\log g \ge 4.8$, which would have been ruled as
non-members based on the \ion{Na}{1}~8190 EW membership cut.

The model does not include Segue~2.  Therefore, model stars that
passed the membership cuts are called contaminants.  We found
\nbesanconcontaminantsraw\ contaminants in the 10~deg$^2$ that we
sampled from the Besan{\c c}on model.  This number needed to be scaled
down by the actual area observed.  To calculate the effective area
observed, we counted the number ($N_{\rm catalog}$) of objects in
0.22~deg$^2$ of the photometric catalog with $14.0 < i_0 < 20.5$ and
$-1.0 < (g-i)_0 < 2.0$.  We also counted the number ($N_{\rm obs}$) of
stars actually observed in the same magnitude and color range, reduced
by the number of spectroscopically confirmed galaxies and members of
Segue~2.  We scaled the number of Besan{\c con} contaminants by the
ratios $(0.22~{\rm deg}^2 / 10~{\rm deg}^2) \times (N_{\rm obs}/N_{\rm
  catalog}$).  The result was \nbesanconcontaminants\ contaminants.
The red dotted histogram in Figure~\ref{fig:vhist} also shows the
velocity distribution of model stars that passed all membership cuts
except radial velocity.  The histogram has also been scaled to reflect
the expected level of contamination for the effective area of sky
observed.

Contaminants could affect the measured velocity dispersion.
Specifically, contaminants would cause an erroneous measurement of a
resolved velocity dispersion or cause the upper limit on the velocity
dispersion to be higher.  Therefore, contaminants do not affect our
conclusion that we cannot resolve the velocity dispersion of Segue~2.
Contaminants could also affect the chemical abundances because those
measurements assumed that all stars were at the distance of Segue~2.
Applied to other stars, the measurements would be meaningless.  Of the
member stars for which we recovered chemical abundances, none of them
are nonsensical, which may indicate that these \nfehmember\ stars are
all true members.

\subsubsection{Comparison to \citeauthor*{bel09}}

Our spectroscopic member sample includes 21 red giants and subgiants,
4 horizontal branch stars, and 1 RR~Lyrae star.  The radial velocity
measurements of all of the stars except the RR Lyrae star contributed
to the measurement of the velocity dispersion of Segue~2.  This sample
size is an improvement over \citeauthor*{bel09}'s spectroscopic
sample, which included 5 red giants, 3 subgiants, and 3 horizontal
branch stars.  Only the red giants contributed to their determination
of the velocity dispersion.  Table~\ref{tab:bel09} shows the
membership classification for the 13 stars in common between our two
samples.  Of these stars, both we and \citeauthor*{bel09} classified
10 as members.  We classified the remaining three stars as
non-members, whereas \citeauthor*{bel09} identified them as part of
the putative stellar stream.  \citeauthor*{bel09} observed one red
giant member that we did not observe.  We do not disagree on the
membership of any Segue~2 star.


\section{Dynamical Properties}
\label{sec:dynamics}

\begin{deluxetable*}{lcc}
\tablewidth{0pt}
\tablecolumns{3}
\tablecaption{Properties of Segue 2\label{tab:parameters}}
\tablehead{\colhead{Property} & \colhead{Symbol} & \colhead{Value}}
\startdata
\cutinhead{Photometry}
Heliocentric distance\tablenotemark{a}       & $D$                            & $35 \pm 2$~kpc \\
Galactocentric distance                      & $D_{\rm GC}$                   & 41 kpc \\
Luminosity\tablenotemark{a}                  & $L_V$                          & $900 \pm 200~L_{\sun}$ \\
Stellar mass\tablenotemark{b}                & $M_*$                          & $1000 \pm 300~M_{\sun}$ \\
Effective (projected half-light) radius\tablenotemark{a} & $R_e$                          & $3.4' \pm 0.2' = 34 \pm 3$~pc \\
3-D (deprojected) half-light radius          & $R_{1/2}$                      & $46 \pm 3$~pc \\
\cutinhead{Dynamics}
Heliocentric mean radial velocity            & $\langle v_{\rm helio} \rangle$ & $-40.2 \pm 0.9$~km~s$^{-1}$ \\
Galactocentric mean radial velocity          & $\langle v_{\rm GC} \rangle$ & $+40.2$~km~s$^{-1}$ \\
Line-of-sight velocity dispersion\tablenotemark{c}            & $\sigma_v$                     & $<2.2~(2.6)$~km~s$^{-1}$ \\
Mass within half-light radius\tablenotemark{c,d} & $M_{1/2}$                    & $<1.5~(2.1) \times 10^5~M_{\sun}$ \\
Mass-to-light ratio within half-light radius\tablenotemark{c,d} & $(M/L_V)_{1/2}$ & $<360~(500)~M_{\sun}/L_{\sun}$ \\
Dynamical-to-stellar mass ratio within half-light radius\tablenotemark{b,c,d} & $(M_{\rm dyn}/M_*)_{1/2}$ & $<300~(410)$ \\
Average density within the half-light radius\tablenotemark{c,d} & $\langle\rho_{1/2}\rangle$ & $<0.4~(0.5)~M_{\sun}~{\rm pc}^{-3}$ \\
\cutinhead{Metallicity}
Mean metallicity\tablenotemark{e}            & $\langle {\rm [Fe/H]} \rangle$ & $-2.22 \pm 0.13$ \\
Standard deviation                           & $\sigma({\rm [Fe/H]})$         & $0.43$ \\
Median metallicity                           & med([Fe/H])                    & $-2.25$ \\
Median absolute deviation                    & mad([Fe/H])                    & $0.27$ \\
Interquartile range                          & IQR([Fe/H])                    & $0.49$ \\
\enddata
\tablerefs{a: \citeauthor*{bel09}.}
\tablenotetext{b}{Assuming that $M_*/L_V = 1.2$.}
\tablenotetext{c}{Upper limits given as 90\% (95\%) C.L.}
\tablenotetext{d}{Using the formula $M_{1/2} = 4 G^{-1} R_e \sigma_v^2$ \protect \citep{wol10}.}
\tablenotetext{e}{Weighted by inverse variance, following \protect \citet{kir11b}.}
\end{deluxetable*}

We estimated $\langle v_{\rm helio} \rangle$ and $\sigma_v$ for
Segue~2 using maximum likelihood statistics and a Monte Carlo Markov
chain (MCMC)\@.  We maximized the logarithm of the likelihood ($L$)
that the given values of $\langle v_{\rm helio} \rangle$ and
$\sigma_v$ described the observed velocity distribution, including the
uncertainty estimates for individual stars.

\begin{eqnarray}
\nonumber \log L &=& \frac{N \log(2 \pi)}{2} + \frac{1}{2} \sum_i^N \left(\log((\delta v_r)_i^2 + \sigma_v^2\right) \\
& & + \frac{1}{2} \sum_i^N \left(\frac{((v_{\rm helio})_i - \langle v_{\rm helio} \rangle)^2}{(\delta v_r)_i^2 + \sigma_v^2}\right)
\end{eqnarray}

\noindent See \citet{wal06} for details regarding this method of
measuring velocity dispersions.  For initial guesses, we started with
\citeauthor*{bel09}'s measurements: $\langle v_{\rm helio} \rangle =
-39.2$~km~s$^{-1}$ and $\sigma_v = 3.4$~km~s$^{-1}$.  Then, we
explored the parameter space with a Metropolis-Hastings implementation
of an MCMC\@.  The length of the chain was $10^8$ trials.

\begin{figure}[t!]
\centering
\includegraphics[width=\columnwidth]{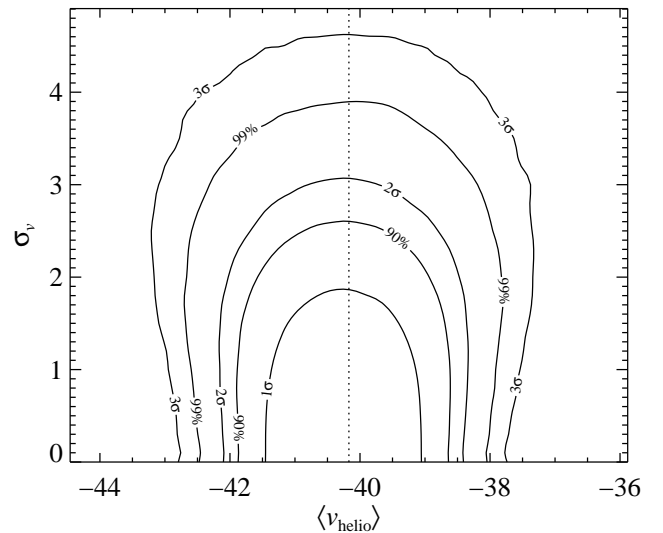}
\caption{Probability contours for the mean velocity and velocity
  dispersion of Segue~2.  The contours show the $1\sigma$ (68.3\%),
  90\%, $2\sigma$ (95.4\%), 99\%, and $3\sigma$ (99.7\%) confidence
  levels.  The dotted line shows the maximum likelihood value of
  $\langle v_{\rm helio} \rangle$.\label{fig:vcontour}}
\end{figure}

\begin{figure}[t!]
\centering
\includegraphics[width=\columnwidth]{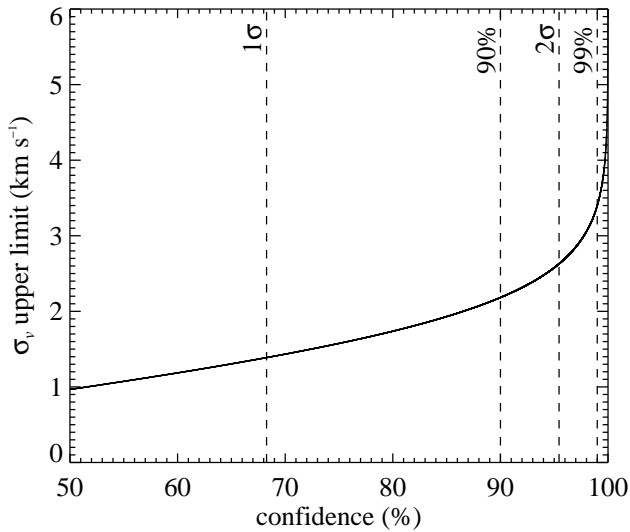}
\caption{Upper limit on the velocity dispersion as a function of
  confidence level.\label{fig:vconfidence}}
\end{figure}

The probability distribution from the MCMC is shown in
Figure~\ref{fig:vcontour}.  Although $\langle v_{\rm helio} \rangle$
is well constrained ($\meanv \pm \meanverr$~km~s$^{-1}$), $\sigma_v$
could not be resolved.  The probability increases toward zero.
Therefore, we could measure only an upper limit for $\sigma_v$.
Figure~\ref{fig:vconfidence} shows the constraint on $\sigma_v$,
marginalized over $\langle v_{\rm helio} \rangle$.  The upper limit on
$\sigma_v$ at the 90\% (95\%) confidence level is $\sigma_v <
\sigmavlimn$ $(\sigmavlimnf)$~km~s$^{-1}$.

Our measurement of $\langle v_{\rm helio} \rangle$ agrees with
\citeauthor*{bel09} to well within the measurement uncertainty.
However, our measurement of $\sigma_v$ is discrepant with
3.4~km~s$^{-1}$ measurement at 99\% confidence.  We have already shown
our individual radial velocity measurements to be consistent.
Furthermore, we counted as members all but one of the stars that
\citeauthor*{bel09} counted as members.  (We did not observe the other
star.)  The origins of the discrepancy in $\sigma_v$ are the
differences in sample sizes and the estimates of uncertainties on
radial velocity.  We used a method to determine $\delta v_r$ nearly
identical to \citeauthor*{bel09}.  However, their estimate of
$\delta_{\rm sys} v_r = 0.35$~km~s$^{-1}$ was based on low-S/N spectra
in the distant Leo~V ultra-faint dwarf galaxy \citep{walker09}.  On
average, \citeauthor*{bel09}'s velocity errors were twice as precise
as those of \citet{walker09}.  Therefore, it may be that their
estimate of $\delta_{\rm sys} v_r$ was inappropriate for spectra with
higher S/N\@.  Regardless, errors in velocity measurements or
underestimates of uncertainty would serve only to decrease the
significance of our upper limit on $\sigma_v$.  The fact that our
upper limit---based on a larger sample---is lower than
\citeauthor*{bel09}'s measurement indicates that our velocity
uncertainties are not underestimated.

The measurement of $\sigma_v$ can be affected by unresolved binary
stars, which would artificially inflate the observed velocity
dispersion.  The problem is especially important for ultra-faint dwarf
galaxies, where binary stars could possibly make a stellar system free
of dark matter appear dark matter-dominated \citep{mcc10}.  In the
case of Segue~1, binaries inflate the observed velocity dispersion by
10\% \citep{mar11}.  We compared our radial velocity measurements to
those of \citeauthor*{bel09}.  All of the radial velocities for the 10
member stars in common agree within the uncertainties.  Therefore,
there is no evidence for a significant inflation of the velocity
dispersion by binaries.  Regardless, our upper limit on the velocity
dispersion is already at the limits of the precision of the radial
velocities.  It would be unlikely for binaries to make the velocity
dispersion appear erroneously unresolved.

If Segue~2 is in equilibrium, then its total mass is related to the
square of the velocity dispersion.  \citet{ill76} devised a formula
appropriate for globular clusters, where the mass distribution follows
the light distribution.  Although this formula has been used for dwarf
galaxies \citep[e.g.,][]{sim07}, it typically underestimates the mass
of galaxies heavily dominated by dark matter.  \citet{wol10} developed
a similar formula appropriate for such systems.  It relates the mass
within the half-light radius to the velocity dispersion and the
deprojected half-light radius.  Because the dark matter profile is
unknown, it is not possible to estimate the total mass with any
accuracy, but the mass within the 3-D half-light radius (46~pc) is
well constrained.  

Because we could not resolve the velocity dispersion, our estimate of
the mass within the half-light radius is an upper limit.  At 90\%
(95\%) confidence, that limit is $M_{1/2} < \masslimn$ $(\masslimnf)
\times 10^5~M_{\sun}$.  The limit on the mass-to-light ratio within
the half-light radius is $(M/L_V)_{1/2} < \mllimn$
$(\mllimnf)~M_{\sun}/L_{\sun}$.

Our upper limit on the mass of Segue~2 makes it the least massive
galaxy known.  Segue~2's small mass raises the possibility that it was
not always so small and that it has instead been tidally stripped by
interaction with the MW\@.  The chemical evidence also supports this
scenario (Section~\ref{sec:stripping}).

The mass estimate depends on the dynamical equilibrium of Segue~2.
The intrinsic line-of-sight velocity distribution for a galaxy in
equilibrium should appear symmetric and roughly Gaussian.  Because the
velocities of all of the stars are consistent with $\langle v_{\rm
  helio} \rangle$, we do not know the intrinsic shape of the velocity
distribution.  In the absence of a better sampled velocity
distribution with smaller uncertainties, we can neither identify
evidence for non-equilibrium dynamics nor conclude that the galaxy is
in equilibrium and supported by dispersion.  However, tidal distortion
of the galaxy would tend to inflate the velocity dispersion, not
depress it.  It is also worth mentioning that the ellipticity of
Segue~2 is small ($0.15 \pm 0.1$, \citeauthor*{bel09}), unlike other
galaxies for which large ellipticities probably indicate tidal
stretching and imminent destruction \citep[e.g., Hercules,][]{dea12}.

The MW cannot be presently disrupting Segue~2 at its current location.
Assuming a MW mass interior to Segue~2's Galactocentric distance
($\distgc$~kpc) of $10^{11}~M_{\sun}$, the upper limit on the Roche
radius is $\rocheradius$~kpc, which is well beyond the extent of the
stars.  On the other hand, if we require that the present tidal radius
be twice the 3-D half-light radius, then the mass of Segue~2 would be
only $\masstide~M_{\sun}$, which is a tiny fraction of the known
stellar mass.  Regardless of whether tides affected Segue~2 in the
past, they are not affecting it now.

\begin{figure*}[t!]
\centering
\includegraphics[width=0.9\textwidth]{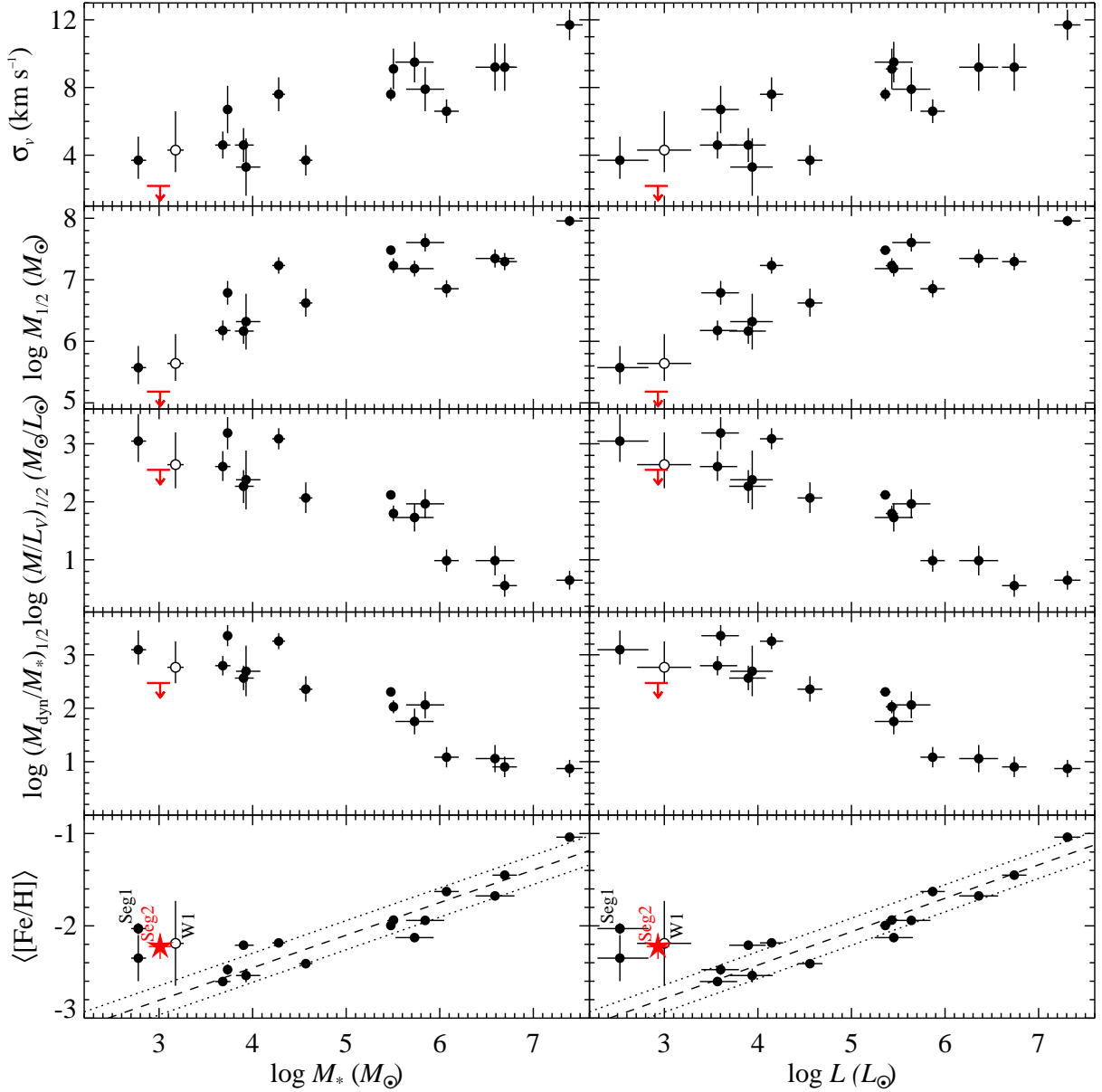}
\caption{Velocity dispersion, mass within the half-light radius,
  mass-to-light ratio within the half-light radius,
  dynamical-to-stellar mass ratio within the half-light radius, and
  average metallicity versus stellar mass (left) and luminosity
  (right) for dwarf galaxies in the Local Group.  Segue~2's upper
  limits at 90\% confidence are represented by red, downward-pointing
  arrows in the top four panels.  Segue~2 is represented by a red star
  in the bottom panel.  Willman~1 is represented by an open circle
  because it may not be in dynamical equilibrium \citep{wil11}.  The
  dashed lines in the bottom panels are the linear regressions taking
  into account errors on both the $x$ and $y$ axes \citep{akr96}.  The
  dotted lines show the rms about the regression.  Dynamical
  quantities ($\sigma_v$, $M_{1/2}$, $(M/L_V)_{1/2}$, and $(M_{\rm
    dyn}/M_*)_{1/2}$) were taken from \citet{mcc12} and references
  therein.  The stellar masses were taken from \citet{woo08} for the
  larger dSphs and \citet[][using the values derived with the
    \citeauthor{kro93}\ \citeyear{kro93} initial mass function]{mar08}
  for the ultra-faint dSphs.  The metallicity for Segue~1 is given
  twice for the two most recent measurements \citep{sim11,var13}.  The
  other metallicities came from \citet[][Willman~1]{wil11} and
  \citet[][other galaxies]{kir11b}.\label{fig:trends}}
\end{figure*}

Figure~\ref{fig:trends} shows various dynamical quantities for MW
satellite galaxies.  They are plotted versus stellar mass, which may
be more relevant to these quantities, and luminosity, which is
directly observable.  Segue~2 is represented as the red,
downward-pointing arrow in the top four panels.  Despite Segue~2's low
mass, the upper limits on its dynamical quantities fall in line with
the envelope defined by other dwarf galaxies.  It seems to be
consistent with a universal relationship between its dynamical mass
and stellar mass.

Figure~\ref{fig:trends} also shows that Segue~2 has the lowest {\it
  total} mass of all of the dwarf galaxies for which the mass has been
estimated.  It is not the galaxy with the lowest luminosity or stellar
mass.  Segue~1 has a lower luminosity and a lower stellar mass than
Segue~2.

\subsection{Possible Stellar Stream}
\label{sec:stream}

\citeauthor*{bel09} identified a possible stellar stream at the same
coordinates and radial velocity as Segue~2.  The criterion for
deciding stream membership rather than galaxy membership was the EW of
the Mg~b triplet, which is an indicator of metallicity.  One of our
membership cuts was that the metallicity of each star, where
measurable, needed to be ${\rm [Fe/H]} < -1$.  The intent of this cut
was to eliminate strong-lined, metal-rich foreground dwarfs as well as
stream members.

\begin{figure}[t!]
\centering
\includegraphics[width=\columnwidth]{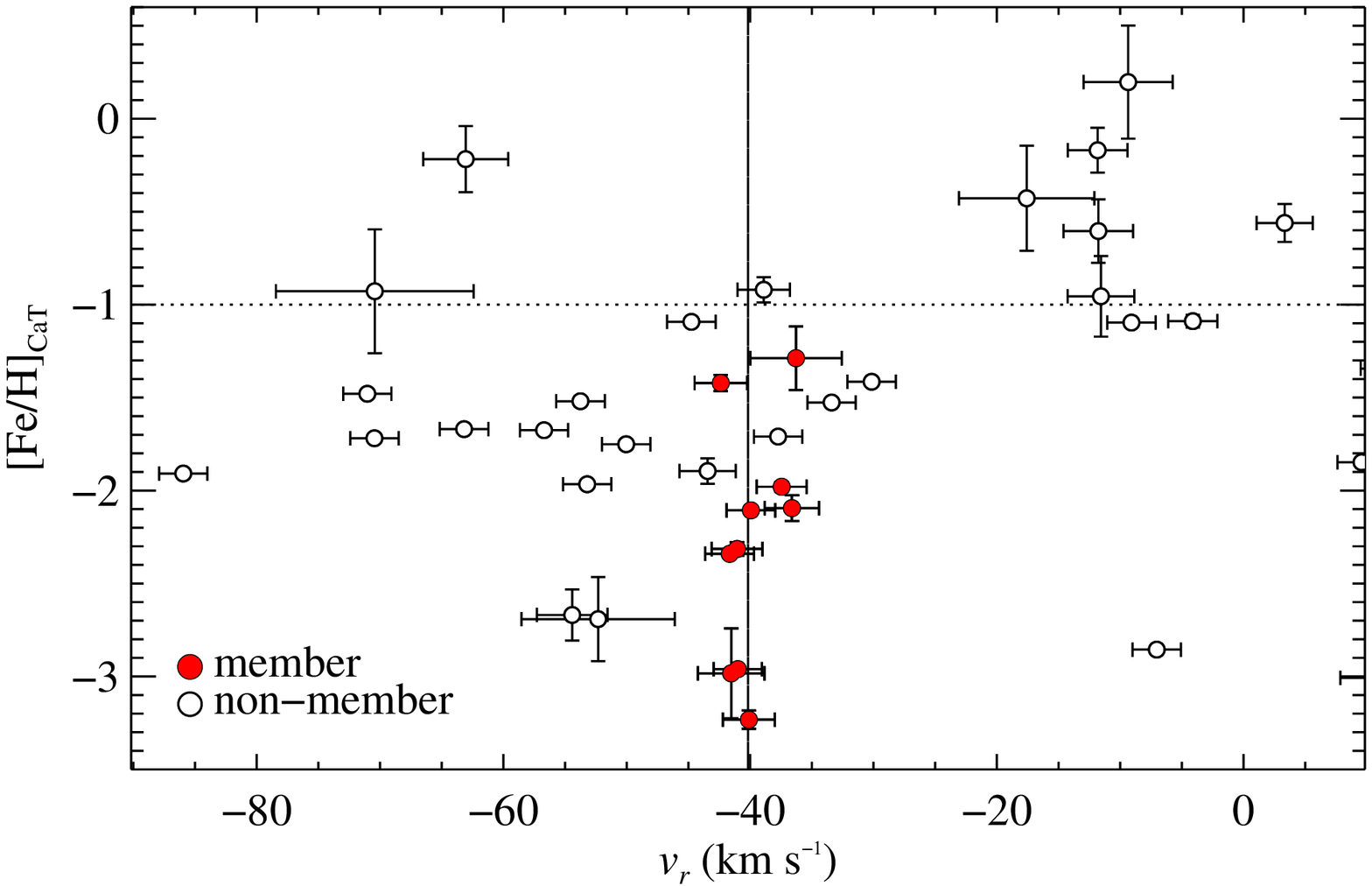}
\caption{Ca triplet metallicity as a function of radial velocity for
  stars that pass all membership cuts except CMD, metallicity, and
  velocity.  Member stars that pass all membership cuts are indicated
  by solid red points.  The vertical dashed line represents $\langle
  v_{\rm helio} \rangle$.  The horizontal dotted line represents the
  metallicity cut for membership, but that cut is based on ${\rm
    [Fe/H]}_{\rm syn}$, not ${\rm [Fe/H]}_{\rm
    CaT}$.\label{fig:fehcatv}}
\end{figure}

Even so, we found no evidence for a tidal stream at the same position
and radial velocity as Segue~2.  Figure~\ref{fig:fehcatv} shows the Ca
triplet metallicity as a function of $v_r$.  This metallicity is
directly related to the EWs of the \ion{Ca}{2}~8542 and 8662
absorption lines.  Therefore, the Ca triplet metallicity is a
diagnostic of stream versus galaxy membership analogous to
\citeauthor*{bel09}'s use of the EW of the Mg~b triplet.  Unlike
\citeauthor*{bel09}, we found no concentration of relatively
strong-lined stars at the same velocity as Segue~2.  Thus, we found no
evidence for a tidal stream.

We observed three of the stars that \citeauthor*{bel09} classified as
stream members.  We ruled all three as non-members on the basis of
radial velocity.  Our radial velocity cut was more restrictive than
that of \citeauthor*{bel09} because they included members based on a
velocity dispersion of $\sigma_v = 3.4$~km~s$^{-1}$.

\citeauthor*{bel09} found evidence for the stream with a spectroscopic
sample covering an area of sky six times larger than our DEIMOS
survey.  The stream is less spatially concentrated than the galaxy.
It is possible that our survey merely missed the stream stars.


\section{Chemical Composition}
\label{sec:chemistry}

We measured metallicities for \nfehmember\ of the \nmember\ member
stars in Segue~2.  We additionally measured at least one of [Mg/Fe],
[Si/Fe], [Ca/Fe], or [Ti/Fe] for \nalphamember\ of those
stars. Table~\ref{tab:parameters} gives some of the metallicity
properties of Segue~2.

\begin{figure}[t!]
\centering
\includegraphics[width=\columnwidth]{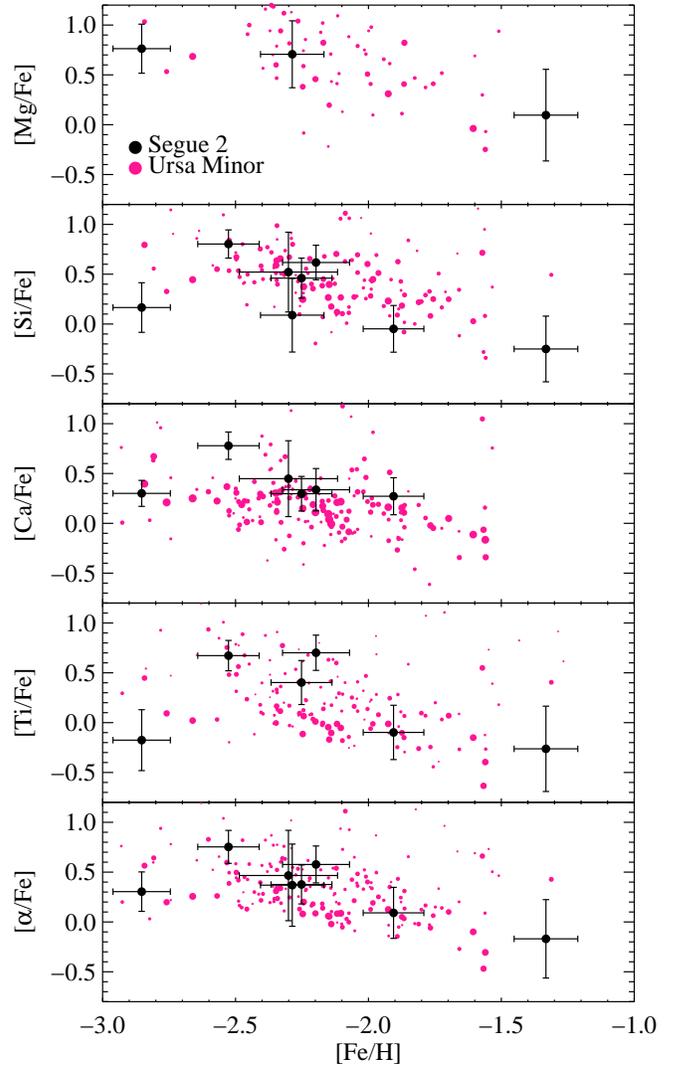}
\caption{Distribution of individual [$\alpha$/Fe] ratios in Segue~2
  (black points) compared to Ursa Minor \citep[magenta points,][]{kir11a},
  a dSph with an average metallicity similar to Segue~2.  The bottom
  panel shows an average of the available [Mg/Fe], [Si/Fe], [Ca/Fe],
  and [Ti/Fe] ratios for each star.  For Ursa Minor, larger points
  represent smaller measurement uncertainties.\label{fig:alpha}}
\end{figure}

Segue~2 has a measurable spread in metallicity ($\sigma({\rm [Fe/H]})
= \fehsigma$).  The spread could be caused by gradual SN enrichment or
inhomogeneous mixing.  Inhomogeneous mixing is especially important
for low-mass, low-metallicity galaxies \citep{gre10,rit12,wis12}.
\citet{fre12} described the effect on chemical abundance patterns of
``one-shot'' chemical enrichment from a single generation of
long-lived stars.  The near instantaneousness of the enrichment would
limit the dispersion in ratios of heavy elements, like [Si/Fe], but
the inhomogeneity could cause large spreads in metallicity indicators,
like [Fe/H]\@.  Figure~\ref{fig:alpha} shows that Segue~2 has
dispersion in both [Fe/H] and heavy element ratios like [Si/Fe]\@.
Therefore, we rule out inhomogeneous mixing as the source of the
metallicity spread.

Instead, the dispersion in metallicity indicates that Segue~2 retained
SN ejecta despite its small mass.  This dispersion stands in contrast
to an ultra-faint globular cluster, like Segue~3 \citep[$\langle {\rm
    [Fe/H]} \rangle = -1.7$, $\sigma({\rm [Fe/H]}) \la
  0.3$,][]{fad11}.  A galaxy with the 90\% C.L.\ upper limit on
Segue~2's velocity dispersion ($\sigmavlimn$~km~s$^{-1}$) could not
have survived the energy input of even a single SN\@.  The SN would
inject $E_{\rm SN} = 8.5 \times 10^{49}$~erg of kinetic energy into
the galaxy \citep{tho98}.  The SN would unbind $M_{\rm ej} = E_{\rm
  ej} / (6 \sigma_v^2) = 1.5 \times 10^5~M_{\sun}$ of gas from the
galaxy.  That mass is 150 times larger than the present stellar mass.
The galaxy must have had much more mass at the time of star formation
than its present stellar mass.  The source of this mass could be dark
matter or stars that were part of the galaxy before any possible tidal
stripping (Section~\ref{sec:stripping}).

The mean metallicity of Segue~2 is $\langle {\rm [Fe/H]} \rangle =
\fehmean \pm \fehmeanerr$, almost the same as the more luminous MW
satellite Ursa Minor \citep[$\langle {\rm [Fe/H]} \rangle = -2.13 \pm
  0.01$,][]{kir11b}.  The similarity in metallicity is notable because
Segue~2 is 330 times less luminous than Ursa Minor.  According to the
universal LZR for MW dSphs \citep{kir11b}, an intact galaxy with the
luminosity of Segue~2 should have a mean metallicity of $\langle {\rm
  [Fe/H]} \rangle = \lzrfeh$.  The intrinsic $1\sigma$ scatter in the
relation at fixed luminosity is 0.16~dex.  Therefore, neither
intrinsic scatter nor the error on the mean metallicity brings Segue~2
into agreement with the LZR\@.  The bottom panels of
Figure~\ref{fig:trends} show the discrepancy, expressed in terms of
luminosity and stellar mass.

The deviation from the LZR is statistically highly significant.  We
sampled \nfehmember\ metallicities from a probability distribution
based on the closed box model of chemical evolution: $P({\rm [Fe/H]})
= (\ln 10/p)10^{\rm [Fe/H]} \exp (-10^{\rm [Fe/H]}/p)$
\citep{lyn75,pag97}.  The effective yield, $p$, is related to the
average metallicity by $\langle {\rm [Fe/H]} \rangle = \log p -
0.251$.  We repeated this sampling for $10^6$ trials.  For each trial,
we sampled the luminosity of Segue~2 from a Gaussian distribution with
a mean of $L = 900~L_{\sun}$ and a width of $200~L_{\sun}$.  We chose
the mean metallicity of the distribution according to the LZR\@.  The
average metallicity of the \nfehmember\ stars met or exceeded the
observed average metallicity of $\langle {\rm [Fe/H]} \rangle =
\fehmean$ in \clzrchance\ trials.  We also replicated the exercise
assuming that the true mean metallicity is 0.16~dex higher than the
LZR to account for the $1\sigma$ intrinsic scatter.  In this case,
\clzrchancedev\ trials out of $10^6$ had $\langle {\rm [Fe/H]} \rangle
\ge \fehmean$.  Therefore, Segue~2 is a significant outlier from the
LZR defined by more luminous dSphs.  Its probability of conforming to
the LZR is at most \lzrchancedev\%.

The detailed abundance pattern of Segue~2 is also similar to Ursa
Minor, measured by \citet{kir11a}.  Figure~\ref{fig:alpha} shows the
comparison.  The distributions are virtually indistinguishable.  The
similar offset and slope of the [$\alpha$/Fe] vs.\ [Fe/H] relation
indicates a similar star formation history for Ursa Minor and Segue~2.
Both galaxies seem to have had very low-level star formation with
declining star formation rates (SFRs) for most of their lives.  The
weak star formation resulted in few Type~II SNe, but the extent of
star formation allowed the later generations of stars to incorporate
the $\alpha$-poor, Fe-rich ejecta from Type~Ia SNe, which were delayed
relative to Type~II SNe.  As a result, the [$\alpha$/Fe] ratios
declined over time as [Fe/H] increased.  The slopes in
Figure~\ref{fig:alpha} for both Segue~2 and Ursa Minor are close to
$-1$, which indicates that the metallicity evolution was driven almost
entirely by Fe and not the $\alpha$ elements.  This can happen only
when the SFR is so low that the frequency of Type~Ia SNe completely
dominates over Type~II SNe after the first generation of stars.
Furthermore, we can deduce that the onset of Type~Ia SNe occurred at
low metallicity in both galaxies (${\rm [Fe/H]} < -2.5$) because a
plateau of [$\alpha$/Fe] is not detectable at higher metallicities.

The decline of [$\alpha$/Fe] with increasing metallicity shows that
star formation in Segue~2 lasted for at least tens of Myr.  The
minimum delay time for a Type~Ia SN is not well constrained, but it
must be at least as long as the lifetime of a star with an initial
mass of $8~M_{\sun}$, the maximum mass of a star that does not explode
as a Type~II SN\@.  That lifetime is about 30~Myr \citep{mat86}.  In
reality, the galaxy needed to sustain several explosions of Type~Ia
SNe to achieve a steady decline in [$\alpha$/Fe]\@.  As a result, the
star formation lifetime is almost certainly at least several times
30~Myr.

Most other ultra-faint satellites share these properties.
\citet{var13} showed that [$\alpha$/Fe] ratios decline as a function
of [Fe/H] in six ultra-faint dwarfs.  Segue~1 and Ursa Major~II are
possible exceptions.  Type~Ia SN enrichment seems to be a nearly
universal phenomenon for dwarf galaxies, including ultra-faint dwarfs.
In contrast, the chemical abundance patterns of almost all globular
clusters do not show evidence for Type~Ia SN enrichment.

The similarity of chemical properties, especially mean metallicity, to
a more luminous satellite leads to two possible explanations for the
origin of Segue~2.  It could be that Segue~2 is a tidally stripped
remnant \citep[see][]{lok12} of a larger satellite.  On the other
hand, it is possible that galaxy formation has a lower bound in
metallicity.

\subsection{The Tidal Stripping Scenario}
\label{sec:stripping}

The similarity of the chemical properties of Ursa Minor and Segue~2
and the deviation from the LZR might indicate that Segue~2 is the
remnant of a galaxy that has been tidally stripped by the MW\@.  In
this scenario, stars would have been removed from Segue~2 as it fell
into the MW's gravitational potential.  It would lose stellar mass and
luminosity as it dissolved, but its stars' chemical properties would
not have changed.  Therefore, it would move to the left in the bottom
panels of Figure~\ref{fig:trends}.  For example, suppose that Segue~2
was a dSph identical to Ursa Minor before it fell into the MW\@.  Its
initial luminosity would be $2.8 \times 10^5~L_{\sun}$ \citep{irw95}.
If it lost 99.7\% of its stellar mass, it would appear as
\citeauthor*{bel09} and we have observed Segue~2.

Tidal stripping of some dwarf galaxies certainly happens.  It can be
dramatic, as is the case for the Sagittarius dSph.  Sagittarius has
tidal tails that span the entire sky \citep{iba02,maj03}, and it
likely deposited many of the MW halo's globular clusters in the
process of dissolution \citep{law10}.  Tidal stripping can also be
subtle and result only in stellar overdensities with cold velocity
dispersions \citep{sch09}.  If the MW is tidally stripping Segue~2,
then it falls into the latter category of subtle disruption, although
\citeauthor*{bel09} suggested that Segue~2 was deposited into the MW
halo by a larger host, just as Sagittarius has deposited globular
clusters.

\citet{pen08a} showed from $N$-body simulations that the $M/L$ will
increase as dwarf galaxies become tidally stripped.  They assumed that
the stars follow cored \citet{kin62} profiles and that the dark matter
follows a cusped NFW \citep{nav97} profile.  In this case, the galaxy
would need to lose nearly all of its mass in both dark matter and
stars in order to end up like Segue~2.  More than 90\% of the dark
matter needs to be stripped before the stars are affected.  In order
to lose 99.7\% of its stellar mass, the galaxy would have to lose
$>99.9\%$ of its dark matter \citep{pen08b}.  However, the dark matter
subhalos of dwarf galaxies may not follow NFW cusps
\citep[i.a.,][]{deb10}, in which case both the stars and dark matter
would be more fragile.  In the case of a cored dark matter profile,
Segue~2 could not have been stripped to its present state without
being completely disrupted.  In the case of a cusped halo, Segue~2
could be the whittled center of a galaxy once the size of Ursa Minor.
Segue~2 may also have its own yet-undetected, low-surface brightness,
low-metallicity tidal tails.

The properties of Segue~2 are mostly consistent with the tidal
stripping scenario except that its half-light radius is very small.
\citet{pen08b} showed that intense tidal stripping of 99\% to 99.9\%
of the stellar mass of a dSph decreases the surface brightness by a
factor of $\sim 300$, increases the mass-to-light ratio by a factor of
$\sim 10$, decreases the velocity dispersion by a factor of $\sim 6$,
and decreases the half-light radius by a factor of $\sim 1.5$.  All of
these except the half-light radius are roughly consistent with
transforming Ursa Minor into Segue~2 via tidal stripping.  The
observed half-light radius of Segue~2 (34~pc) is a factor of 3--4 too
small.  However, \citet{pen08b} considered only six models of dwarf
satellite galaxies.  A larger parameter space might include galaxies
whose half-light radii are reduced enough to explain Segue~2 within
the tidal stripping scenario.

In the tidal disruption scenario, the upper limits on the dynamical
quantities that we calculated from the upper limit on the velocity
dispersion may be incorrect.  The upper limit on the mass estimate
assumed that the galaxy is in dynamical equilibrium and is supported
by velocity dispersion.  If Segue~2 is instead an unbound tidal stream
or loosely bound galaxy, then the mass estimate is unreliable.
Nonetheless, the upper limits for the dynamical quantities for Segue~2
are consistent with the same relationships between dynamical
quantities and luminosity or stellar mass as other dwarf galaxies
(Figure~\ref{fig:trends}).

\subsection{A Metallicity Floor for Galaxy Formation}
\label{sec:floor}

An alternative explanation for the chemical composition and low
luminosity of Segue~2 is that the stellar content of the galaxy is
bound and largely unaffected by tidal stripping.  Instead, the galaxy
formed only $\mstar~M_{\sun}$ of stars over its entire lifetime.  In
this case, the comparatively high metallicity of the galaxy needs to
be explained.

Segue~2 is not alone in lying above the LZR\@.  The ultra-faint
satellites Segue~1 and Willman~1 also might be more metal-rich than
their luminosities would suggest from an extrapolation of the LZR\@.
Segue~1 contains an extremely metal-poor star \citep[${\rm [Fe/H]} =
  -3.3 \pm 0.2$,][]{geh09}.  However, more recent spectroscopy
\citep{sim11} of more stars found a mean metallicity of $\langle {\rm
  [Fe/H]} \rangle = -2.38 \pm 0.37$.  \citet{var13} independently
measured $\langle {\rm [Fe/H]} \rangle = -2.03 \pm 0.06$ from the same
spectra of Segue~1 stars.  \citet{nor10} also derived $\langle {\rm
  [Fe/H]} \rangle = -2.7 \pm 0.4$ from Ca~H and K absorption rather
than \ion{Fe}{1} lines.  For its luminosity \citep{mar08}, Segue~1
should have a mean metallicity of $\langle {\rm [Fe/H]} \rangle =
\lzrfehsegone$.  \citeauthor{sim11}'s measurements are marginally
consistent with the LZR, taking into account the error on the mean and
the intrinsic scatter in the relation.  On the other hand, Segue~1 is
a highly significant outlier based on \citeauthor{var13}'s
measurements.  Willman~1 is another satellite that might lie above the
LZR\@.  Unfortunately, there are only two known red giants in the
galaxy, and their average metallicity is $\langle {\rm [Fe/H]} \rangle
= -2.19 \pm 0.46$ \citep{wil11}.  Based on its luminosity, it should
have $\langle {\rm [Fe/H]} \rangle = \lzrfehwilone$.  Furthermore, the
galaxian nature of Willman~1 is uncertain.  The velocity distribution
of its stars is irregular and non-Gaussian \citep{wil11}.  It may not
be in dynamical equilibrium, and it could be a tidally stripped
remnant, such as we proposed for Segue~2 in
Section~\ref{sec:stripping}.

It is not practical to draw strong conclusions from only three
galaxies that disobey the LZR.  However, Segue~1, Segue~2, and
Willman~1 together invoke the suggestion of a metallicity floor for
galaxy formation.  \citet{sim07} first noticed that the ultra-faint
galaxies lie slightly above the LZR\@.  However, they used a
metallicity indicator that has since been shown to be unreliable for
very metal-poor stars.  \citet{kir11a} recomputed average
metallicities based on \ion{Fe}{1} lines, and they noticed a subtle
change in the slope of the LZR around $10^5~L_{\sun}$ in the sense
that the ultra-faint galaxies lay slightly above the LZR extrapolated
from higher luminosities.

There is a theoretical reason to expect a metallicity floor for galaxy
formation.  A single pair instability SN can bring the metallicity of
the interstellar medium of a galaxy to about one thousandth of the
solar value ($10^{-3}~Z_{\sun}$) and mark the transition from
Population~III to II star formation \citep{wis12}.  Although one
thousandth of the solar metallicity is not far from the metallicity
floor that we observed (${\rm [Fe/H]} \approx -2.5$), some dwarf
galaxies have stars with metallicities a factor of ten below that
floor \citep[e.g.,][]{fre10,taf10}.  As \citeauthor{wis12}\ pointed
out, the transition from Population~III to II may not be reflected in
a hard boundary in metallicity.  But it could be reflected in a floor
for the average metallicities of galaxies.  Damped Lyman $\alpha$
systems (DLAs) at redshifts of $2.5 < z < 5$, which are presumably
nascent galaxies, do not exhibit metallicities below $10^{-3}$ of the
solar metallicity \citep{pro03,raf12}.  It is conceivable that
galaxies do not form unless they achieve this threshold metallicity.
However, DLAs have much larger gas masses than the stellar mass of
Segue~2.

We reiterate that it is unwise to make much of the sparse sampling of
the LZR at $L < 10^4 L_{\sun}$.  It would be wise to expand the sample
sizes in Segue~1, Segue~2, and Willman~1 before we conclude that the
linearity of the LZR disappears below a few thousand $L_{\sun}$\@.
Nonetheless, it is interesting to consider the bottom panels of
Figure~\ref{fig:trends} together with the metallicity floor for
DLAs\@.


\section{Relevance to Dark Matter Physics}
\label{sec:dark_matter_physics}

The two models presented in the previous section for the origin of
Segue~2's high metallicity given its luminosity have very different
implications for our understanding of galaxy formation and cold dark
matter (CDM) models.  In this section, we address Segue~2's place in
the $\Lambda$CDM paradigm.

In Section~\ref{sec:stripping}, we suggested that Segue~2 may be the
remnant of a galaxy that was previously 100--1000 times more luminous
than it is now.  Furthermore, we inferred from its retention of
supernova ejecta that it was once hosted by a substantial dark matter
halo.  If so, then Segue~2 is the first known galaxy to have shed its
dark matter halo without being completely disrupted.  To be consistent
with a progenitor similar to Ursa Minor, which provides a good match
for the average metallicity and abundance ratios, Segue~2 would have
once been hosted by a halo with a maximum circular velocity of $v_{\rm
  max} \approx 20-25$~km~s$^{-1}$, corresponding to a virial mass of
$M_{\rm vir} \approx (1-3) \times 10^{9}~M_{\sun}$ \citep{boy12}.

By number, such objects should be rare---but not extremely rare---in
the MW at $z=0$.  The Aquarius simulations \citep{spr08} have between
60 and 110 distinct objects within 300~kpc of the halo center at $z=0$
that were once at least this massive \citep[][Table~1]{boy12}.
However, most of these reside at large distances from their halo's
center and have apocenter-to-pericenter ratios around $6:1$, which is
not conducive to strong tidal stripping.  The ratio needed to achieve
stripping of 99.7\% of the stars (to transform an Ursa Minor-like dSph
into Segue~2) would have to be about $50:1$ \citep{pen08b}.  Only a
small fraction of subhalos have orbital eccentricities that large.
Those subhalos would spend most of their lives at large Galactocentric
distances.  As a result, galaxies similar to Segue~2 would be
difficult to detect, and Segue~2 would likely be one of the closest
such objects to the MW's center.  Furthermore, extremely tidally
stripped galaxies likely would not survive their next 1--2 pericentric
passages, which would make Segue~2 a transient phenomenon that will
survive for only 2--3~Gyr.  Proper motion studies of Segue~2 would be
valuable in establishing whether its orbit is consistent with the
tidal stripping hypothesis.

Even in the absence of proper motions, the Galactic coordinates
($l=149\arcdeg$, $b=-38\arcdeg$) and Galactocentric distance
(\distgc~kpc) of Segue~2 exclude it from being a member of the MW's
``disk of satellites'' suggested by \citet{met07}.  The disk is $\sim
20$~kpc thick, and its pole points toward $l=158\arcdeg$,
$b=-12\arcdeg$.  \citeauthor{met07} claimed that the existence of the
plane was inconsistent with CDM structure formation.  Segue~2's
location does not invalidate the plane of satellites because many of
the MW's satellites, including the most luminous satellites, do lie in
the plane.  However, Segue~2 could not be a tidal dwarf galaxy that
came from the merger event that \citeauthor{met07} proposed to explain
the orientation of the plane-aligned satellites.

In Section~\ref{sec:floor}, we presented an alternate scenario,
wherein Segue~2 is not a heavily-stripped remnant of a more massive
galaxy but instead a relatively undisturbed dwarf galaxy whose
properties indicate the existence of a metallicity floor in galaxy
formation.  The implications for $\Lambda$CDM galaxy formation in this
scenario are very different from the tidal stripping scenario.
Segue~2 would have been born in a very low-mass dark matter halo of
$v_{\rm max} \la 8$~km~s$^{-1}$.  The MW is predicted to host at least
1000 such dark matter subhalos at present in CDM models
\citep{die08,spr08}.  Either Segue~2 would be the first of a vast
class of new galaxies to be discovered with very low luminosities and
very low dark matter content, or it would have to represent a rare
case of a dark matter halo that is typically too small to host a
galaxy but, for some reason, managed to form a small number of stars
over at least 100~Myr.

The presence of stars in a subhalo as small as $v_{\rm max} =
8$~km~s$^{-1}$ is especially remarkable because the threshold for
atomic cooling to form stars is $v_{\rm max} > 17$~km~s$^{-1}$
\citep{ree77}.  Expressed in terms of virial temperature, Segue~2 has
$T_{\rm vir} < 2400$~K, whereas the atomic hydrogen cooling threshold
is $10^4$~K.  As a result, the gas in Segue~2 would have had to
undergo molecular (H$_2$) cooling in order to form stars.  If the mass
of the subhalo when it formed stars was as low as it is today, then
Segue~2 would be the only known galaxy that must have experienced only
molecular cooling without atomic cooling.

If Segue~2's dark matter halo has not been heavily affected by
Galactic tides, then its existence might provide an interesting
constraint on Warm Dark Matter (WDM) models because it has a low halo
mass ($v_{\rm max} < 8$~km~s$^{-1}$).  The formation of halos with
$v_{\rm max} \sim 8$~km~s$^{-1}$ should be strongly suppressed for
thermal WDM particles with $m_{\rm WDM} \sim 4$~keV
\citep{sch13,ang13}, comparable to published lower limits based on the
Lyman~$\alpha$ forest \citep{vie09}.  The existence of galaxies in
very low mass dark matter halos would favor models with significant
small-scale power (e.g., CDM), and counts of such objects may provide
the strongest lower limits on the thermal mass of dark matter
particles \citep{pol11}.  However, more work is needed to clarify the
extent to which halo formation is suppressed below the associated
filtering scale in the power spectrum \citep{ang13}.

\section{Summary}
\label{sec:summary}

\citet{bel09} discovered the Segue~2 ultra-faint dwarf galaxy in SEGUE
photometry.  From the line-of-sight stellar velocity dispersion
measured with MMT/Hectospec, they determined that its mass has a much
larger fraction of dark matter than of stars.  They also found
evidence for a tidal stream at the same position and radial velocity
as Segue~2.

We observed Segue~2 with ten Keck/DEIMOS slitmasks in order to
enlarge the sample of radial velocities, refine the mass estimate, and
add metallicity and detailed chemical abundances to the available
data.  We measured radial velocities by cross-correlation with
template spectra, and we estimated velocity uncertainties by Monte
Carlo resampling of the spectra based on a noise model of the full
spectrum.  We measured Mg, Si, Ca, Ti, and Fe abundances by comparing
the observed spectra to synthetic spectra using $\chi^2$ minimization.

With our threefold increase in spectroscopic sample size, we could not
confirm \citeauthor*{bel09}'s measurements of velocity dispersion
($3.4^{+2.5}_{-1.2}$~km~s$^{-1}$) and mass ($5.5^{+10.5}_{-3.1} \times
10^5~M_{\sun}$).  Instead, we placed an upper limit on the stellar
line-of-sight velocity dispersion of $\sigma_v < \sigmavlimn$
$(\sigmavlimnf)$~km~s$^{-1}$ with 90\% (95\%) confidence.  The
inferred limit on the mass within the half-light radius is $M_{1/2} <
\masslimn$ $(\masslimnf) \times 10^5~M_{\sun}$.

We found a dispersion in [Fe/H] and a decline of [$\alpha$/Fe] as a
function of [Fe/H]\@.  These two chemical properties establish that
Segue~2 retained SN ejecta and that star formation lasted for at least
several generations of Type~Ia SNe (at least 100~Myr).  We also
determined that the average metallicity of Segue~2 places it more
metal-rich than the LZR defined by the more luminous MW satellite
galaxies.  The [$\alpha$/Fe] ratios as a function of metallicity are
indistinguishable from the more luminous dSph Ursa Minor.

Taken together, the dynamical and chemical characteristics of Segue~2
point to two possible scenarios for its formation.  Segue~2 could be
the barest remnant of a galaxy that once had a luminosity of $2 \times
10^5~L_{\sun}$.  Gravitational interaction with the MW's gravitational
potential removed nearly all of of its stars and dark matter halo,
leaving only the dense center of the galaxy.  On the other hand,
Segue~2 could have formed with its present stellar mass and
metallicity.  This scenario has been proposed for the ultra-faint
galaxy Segue~1 \citep{geh09,sim11}.  The deviation from the LZR might
indicate a metallicity floor for galaxy formation, which is evocative
of the metallicity floor observed for damped Lyman $\alpha$ systems at
high redshift \citep{pro03,raf12}.

Overall, the tidal stripping scenario seems more plausible.  The
metallicity floor scenario must overcome the fact that dSphs contain
extremely metal-poor stars and that there are ultra-faint galaxies,
like Coma Berenices \citep[$\langle {\rm [Fe/H]} \rangle = -2.60 \pm
  0.05$,][]{kir11a}, with metallicities below that of Segue~2.  On the
other hand, SDSS has revealed copious evidence of ongoing tidal
stripping of satellites, and Segue~2 may not have been immune to the
same fate.  The distinguishing characteristic of Segue~2 is that the
MW's tides have likely whittled it down to the least massive galaxy
known.  The tidal stripping scenario still needs to overcome the
discrepancy between the half-light radius of models of tidally
stripped dSphs \citep{pen08b} and the observed half-light radius of
Segue~2.  It would be interesting to explore a larger parameter space
of initial galaxy shapes \citep[such as the recent simulations
  of][]{kaz13} to test whether tidal stripping of a dSph with the
stellar mass of Ursa Minor could produce Segue~2.

On a final note, we reached the limit of measuring a galaxy's velocity
dispersion with DEIMOS\@.  Resolving the dispersion of Segue~2 will
require higher resolution spectroscopy.  High-resolution spectrographs
like Keck/HIRES and Subaru/HDS could possibly measure the velocity
distribution of $\sim 10$ stars in Segue~2 over 3--4 nights.  However,
new surveys like LSST and SkyMapper will discover new tiny galaxies,
some of which may have velocity dispersions too small to be resolved
with medium-resolution, multi-object spectrographs.  These tiny
galaxies and their relevance to dark matter physics make a compelling
case for building high-resolution spectrographs for the next
generation of extremely large telescopes.  The combination of
collecting area and spectral resolution will allow measurements of the
velocities of the fainter, more common stars in Segue~2 and
undiscovered galaxies like it.  Only then can we measure their
dynamical masses.

\acknowledgments We are grateful to the many people who have worked to
make the Keck Telescope and its instruments a reality and to operate
and maintain the Keck Observatory.  The authors wish to extend special
thanks to those of Hawaiian ancestry on whose sacred mountain we are
privileged to be guests.  Without their generous hospitality, none of
the observations presented herein would have been possible.

We thank the anonymous referee for a courteous report on our article
and Josh Simon for a helpful discussion.  We are also grateful to
Shunsaku Horiuchi for confirming that Segue~2 shows no signal in sky
maps from the {\it Fermi} Gamma Ray Space Telescope.  ENK and MBK
acknowledge support from the Southern California Center for Galaxy
Evolution, a multicampus research program funded by the University of
California Office of Research, and partial support from NSF grant
AST-1009973.  JGC thanks NSF grant AST-0908139 for partial support.

{\it Facility:} \facility{Keck:II (DEIMOS)}

\clearpage
\renewcommand{\thetable}{\arabic{table}}
\setcounter{table}{1}
\begin{turnpage}
\tabletypesize{\scriptsize}
\begin{deluxetable*}{lccccccccc}
\tablewidth{0pt}
\tablecolumns{10}
\tablecaption{Target List\label{tab:catalog}}
\tablehead{\colhead{ID} & \colhead{$g_0$} & \colhead{$i_0$} & \colhead{Masks\tablenotemark{a}} & \colhead{S/N} & \colhead{$v_r$} & \colhead{EW(\ion{Na}{1}~8190)} & \colhead{[Fe/H]} & \colhead{Member?} & \colhead{Reason\tablenotemark{b}} \\
\colhead{ } & \colhead{(mag)} & \colhead{(mag)} & \colhead{ } & \colhead{(\AA$^{-1}$)} & \colhead{(km~s$^{-1}$)} & \colhead{(\AA)} & \colhead{(dex)} & \colhead{ } & \colhead{ }}
\startdata
SDSS J021908.97+201948.9 & $23.091 \pm 0.331$ & $19.745 \pm 0.036$ & 1 &  52.2 & $ -65.5 \pm  2.0$ &       \nodata     &     \nodata      & N &       CMD $v_r$  \\
SDSS J021909.11+201115.9 & $21.178 \pm 0.075$ & $20.836 \pm 0.101$ & 1 &  17.4 & $ -43.2 \pm 10.7$ &       \nodata     &     \nodata      & Y &                  \\
SDSS J021909.23+200552.8 & $25.405 \pm 0.426$ & $21.478 \pm 0.184$ & 1 &  17.1 & $ +32.0 \pm  3.0$ &       \nodata     &     \nodata      & N &       CMD $v_r$  \\
SDSS J021909.29+201958.7 & $22.572 \pm 0.219$ & $23.509 \pm 0.721$ & 1 &   0.3 &      \nodata      &       \nodata     &     \nodata      & N &          CMD Bad \\
SDSS J021909.34+200045.5 & $24.447 \pm 0.568$ & $21.909 \pm 0.201$ & 1 &   6.2 &      \nodata      &       \nodata     &     \nodata      & N &          CMD Bad \\
SDSS J021909.53+200056.5 & $21.681 \pm 0.107$ & $19.389 \pm 0.025$ & 1 &  63.4 & $ +35.1 \pm  2.0$ &       \nodata     &     \nodata      & N &       CMD $v_r$  \\
SDSS J021909.84+201122.7 & $22.818 \pm 0.294$ & $20.507 \pm 0.073$ & 1 &   8.2 & $ +28.6 \pm  5.0$ &       \nodata     &     \nodata      & N &       CMD $v_r$  \\
SDSS J021909.97+201254.0 & $20.252 \pm 0.033$ & $19.575 \pm 0.032$ & 2 &  49.4 & $ -42.4 \pm  2.1$ &       \nodata     & $-1.33 \pm 0.12$ & Y &                  \\
SDSS J021910.17+201539.1 & $24.182 \pm 5.490$ & $17.341 \pm 0.013$ & 1 & 110.5 & $ +13.6 \pm  2.0$ &       \nodata     &     \nodata      & N &       CMD $v_r$  \\
SDSS J021910.22+200324.8 & $19.593 \pm 0.021$ & $19.374 \pm 0.024$ & 1 &  61.2 & $ +93.5 \pm  7.6$ &       \nodata     &     \nodata      & N &       CMD $v_r$  \\
\nodata & \nodata & \nodata & \nodata & \nodata & \nodata & \nodata & \nodata & \nodata & \nodata \\
\enddata
\tablerefs{Identifications and photometry from SDSS \protect \citep{aba09}.}
\tablenotetext{a}{Number of DEIMOS masks on which the object was observed.}
\tablenotetext{b}{Reasons for non-membership.  CMD: Location in the color-magnitude diagram.  $v_r$: Inappropriate radial velocity.  Na: Spectrum shows strong \ion{Na}{1}~$\lambda$8190 doublet.  [Fe/H]: The measured metallicity is greater than ${\rm [Fe/H]} = -1.0$.  G: Spectrum shows emission lines or redshifted Ca H and K lines, indicating that the object is a galaxy.  Bad: Spectral quality was insufficient for radial velocity measurement.}
\tablecomments{(This table is available in its entirety in a machine-readable form in the online journal. A portion is shown here for guidance regarding its form and content.)}
\end{deluxetable*}
\clearpage
\end{turnpage}

\end{document}